\documentclass[iop]{emulateapj}

\usepackage{graphicx}
\usepackage{apjfonts}

\newcommand{\domm}[1]{\ifmmode #1\else$#1$\fi}

\newcommand{\azero}{\domm{a_0}}
\newcommand{\Msun}{\domm{M_{\odot}}}
\newcommand{\Rsun}{\domm{R_{\odot}}}

\let\Mchan\Mch

\newcommand{\psec}{\domm{{\rm s}^{-1}}}
\newcommand{\gcc}{\domm{{\rm g\,cm}^{-3}}}

\newcommand{\Md}{\domm{M_{\rm d}}}
\newcommand{\Ma}{\domm{M_{\rm a}}}
\newcommand{\Ra}{\domm{R_{\rm a}}}

\newcommand{\qm}{\domm{q_\mathrm{m}}}
\newcommand{\Omegamax}{\domm{\Omega_{\rm max}}}
\newcommand{\MencTmax}{\domm{M_{\rm enc}(T_{\rm max})}}
\newcommand{\zMencTmax}{\domm{M_{\rm enc}(T^{\rm z}_{\rm max})}}

\newcommand{\rhocrat}{\domm{\rho_{\rm c,d}/\rho_{\rm c,a}}}
\newcommand{\rhoc}{\domm{\rho_{\rm c}}}
\newcommand{\rhocd}{\domm{\rho_{\rm c,d}}}
\newcommand{\rhoca}{\domm{\rho_{\rm c,a}}}
\newcommand{\qrho}{\domm{q_\rho}}

\newcommand{\MMfifty}{\domm{M_{\rm enc}(\frac{1}{2}M_{\rm d})}}
\newcommand{\MMthick}{\domm{\Delta M_{\rm enc}(M_{\rm d})}}
\newcommand{\Mrem}{\domm{M_{\rm ce}}}
\newcommand{\Mdisk}{\domm{M_{\rm disk}}}

\newcommand{\Tmax}{\domm{T_{\rm max}}} 
\newcommand{\zTmax}{\domm{T^{\rm z}_{\rm max}}} 
\newcommand{\rhoTmax}{\domm{\rho(T_{\rm max})}}
\newcommand{\zrhoTmax}{\domm{\rho(T^{\rm z}_{\rm max})}}
\newcommand{\MEthermfifty}{\domm{M_{\rm enc}(\frac{1}{2}E_{\rm th})}}
\newcommand{\MEthermthick}{\domm{\Delta M_{\rm enc}(E_{\rm th})}}
\newcommand{\MencOmax}{\domm{M_{\rm enc}(\Omega_{\rm max})}}
\newcommand{\Omegaorb}{\domm{\Omega_{\rm orb}}}
\newcommand{\MErotfifty}{\domm{M_{\rm enc}(\frac{1}{2}E_{\rm rot})}}
\newcommand{\MErotthick}{\domm{\Delta M_{\rm enc}(E_{\rm rot})}}

\newcommand{\hz}{\domm{h_\mathrm{z}}}
\newcommand{\hxy}{\domm{h_\varpi}}
\newcommand{\rhocrhoct}{\domm{\rho_{\rm c}$/$\rho_{\rm c,a}}}
\newcommand{\Tc}{\domm{T_{\rm c}}}

\newcommand{\Lrat}{\domm{L_{\rm ce}/L_{\rm tot}}}
 
\newcommand{\mrm}{\mathrm}
\newcommand{\rxy}{$\varpi$}	
\newcommand{\fdon}{\domm{f_{\rm d}}}
\newcommand{\fratio}{\domm{(f_{\rm d}/f_{\rm a})_{\rm cc}}}

\newcommand{\eqbegin}{\begin{equation}}
\newcommand{\eqend}{\end{equation}}

\defcitealias{vkercj10}{vKCJ10}
\defcitealias{loreig09}{LIG09}
\newcommand{\citeal}{\citetalias}

\begin{document}

\title{A Parameter-Space Study of Carbon-Oxygen White Dwarf Mergers}

\author{Chenchong Zhu, Philip Chang, Marten H. van Kerkwijk and James Wadsley}
\email{cczhu@astro.utoronto.ca}

\date{\today}

\begin{abstract}
The merger of two carbon-oxygen white dwarfs can lead either to a spectacular transient, stable nuclear burning or a massive, rapidly rotating white dwarf.  Simulations of mergers have shown that the outcome strongly depends on whether the white dwarfs are similar or dissimilar in mass.  In the similar-mass case, both white dwarfs merge fully and the remnant is hot throughout, while in the dissimilar case, the more massive, denser white dwarf remains cold and essentially intact, with the disrupted lower mass one wrapped around it in a hot envelope and disk.
In order to determine what constitutes ``similar in mass'' and more generally how the properties of the merger remnant depend on the input masses, we simulated unsynchronized carbon-oxygen white dwarf mergers for a large range of masses using smoothed-particle hydrodynamics.
We find that the structure of the merger remnant varies smoothly as a function of the ratio of the central densities of the two white dwarfs.  A density ratio of 0.6 approximately separates similar and dissimilar mass mergers.  Confirming previous work, we find that the temperatures of most merger remnants are not high enough to immediately ignite carbon fusion.
During subsequent viscous evolution, however, the interior will likely be compressed and heated as the disk accretes and the remnant spins down.  We find from simple estimates that this evolution can lead to ignition for many remnants.  For similar-mass mergers, this would likely occur under sufficiently degenerate conditions that a thermonuclear runaway would ensue.

\end{abstract}

\maketitle

\section{Introduction}
\label{sec:intro}

A few percent of all white dwarfs (WDs) will eventually merge with another white dwarf.  The outcome of such mergers will depend on the compositions of the WDs involved.  For two helium WDs, a low-mass helium star might result, which would be observed as an sdOB star.  For a helium WD merging with a carbon-oxygen one, a helium giant could form, observable as a hydrogen-deficient giant or R CrB star.  For two carbon-oxygen WDs (CO WDs), the outcome could vary between simply a more massive WD, a carbon-burning star, an explosion, or collapse to a neutron star, depending on whether stable or unstable carbon fusion is ignited, and whether the total mass exceeds the critical mass for pycnonuclear ignition or electron captures (both close to the Chandrasekhar mass \Mchan).  For mergers involving an oxygen-neon WD, the mass will always be high, and explosive demise or transmutation seems inevitable.

The outcome of the merger of two CO WDs is uncertain in part because during the merger temperatures do not become hot enough to ignite significant carbon fusion (e.g., \citealt{loreig09}, \citetalias{loreig09} hereafter), except possibly for masses above $\sim\!0.9\,M_\odot$ \citep{pakm+11,pakm+12}.  Hence, the final fate depends on subsequent evolution, in which differential rotation is dissipated, the remnant disk accretes, and the whole remnant possibly spins down.  Due to these processes, the remnant could be compressed and heated, which, if it happens faster than the thermal timescale, would lead to increased temperatures and thus potentially to ignition.  

So far, efforts have focused on merging binaries with total mass $M>\Mchan$.  The end result of such mergers is believed to be either stable off-center carbon ignition, which would turn the merger remnant into an oxygen-neon WD and possibly eventually result in accretion-induced collapse \citep{saion98}, or slow accretion, which allows the remnant to stay cool and eventually ignite at high central density \citep{yoonpr07}.  Less massive mergers were usually thought to result in more massive, rapidly rotating CO WDs \citep{segrcm97,kube+10}, but more recently it has been realized these might eventually become hot enough to ignite (\citealt{vkercj10}, \citeal{vkercj10} hereafter; \citealt{shen+12,schw+12}).  Indeed, \citeal{vkercj10} argue that type Ia supernovae result generally from mergers of CO WDs with similar masses, independent of whether or not their total mass exceeds \Mchan\ (see below).  For all these studies, the conclusions on whether and where ignition takes place depend critically on the structure of the merger remnnant.

The merging process, and the merger remnant, have been studied quite extensively, mostly using smoothed-particle hydrodynamics (SPH; e.g. \citealt{mona05}).  These simulations have shown that the outcome strongly depends on whether the WDs are similar or dissimilar in mass.  In the similar-mass case, both WDs disrupt fully and the remnant is hot throughout, while in the dissimilar case, the more massive, denser WD remains essentially intact and relatively cold, with the disrupted lower mass one wrapped around it in a hot envelope and disk.  Less clear, however, is what constitutes ``similar-mass,'' and, more generally, how the merger remnant properties depend on the initial conditions.  

In principle, for cold WDs of given composition, the remnant properties should depend mostly on the two WD masses, with a second-order effect due to rotation.  In this paper, we try to determine these dependencies using simulations of WD mergers with the Gasoline SPH code, covering the entire range of possible donor and accretor masses, but limiting ourselves to non-rotating WDs.  Our primary aim is to identify trends between mergers of different masses, both to guide analytical understanding and to help scale other, perhaps more precise simulations.  Here, our hope is that while the results of individual simulations may suffer from uncertainties related to the precise techniques and assumptions used, the trends should be more robust.  We also try to provide sufficient quantitative detail on the properties of merger remnants that it becomes possible to make analytical estimates or construct reasonable numerical approximations without having to run new simulations.

Our work is complementary to the recent surveys of remnant properties by \cite{rask+12} and \cite{dan+12}, in that they focus on different scientific questions (e.g., orbital stability; possible detonation).  In contrast to our work, they assume that the WDs are co-rotating with the orbit.  Whether this is a better assumption than no rotation depends on the strength of tidal dissipation, which unfortunately is not yet known (see \citealt{marsns04,fulll12}).

Our work also is part of a series of numerical studies investigating the viability of sub-Chandrasekhar mass (sub-\Mchan) CO WD mergers producing SNe Ia, as proposed by \citetalias{vkercj10}.  This channel relies on similar-mass mergers producing remnants that are hottest near the center, and on compressional heating by subsequent accretion and/or magnetically mediated spin-down leading to ignition.  The advantages of this channel are that it accounts for the absence of direct evidence for stellar companions, the observed SN Ia rate, and the dependence of SN Ia peak luminosity on the age of the host stellar population (because lower-mass merger constituents take longer to form).  Since pure detonations of sub-\Mchan\ CO WDs produce light curves very similar to observed SNe Ia \citep{shig+92,sim+10}, it also removes the need for imposed deflagration-to-detonation transitions.  Important questions, however, remain, including what fraction of mergers leads to remnants that are hot near the center (in highly degenerate conditions), how the subsequent viscous phase proceeds in detail, whether ignition leads to a detonation, and whether the detonation of a remant that may still rotate and be surrounded by a disk would produce an event similar to an SN Ia.  With our work, we attempt to address the first question.

This paper is organised as follows.  In Section~\ref{sec:gasoline}, we describe the SPH code we used, as well as our initial conditions.  In Section~\ref{sec:results}, we present our results and give trends for a number of pertinent remnant properties.  In Section~\ref{sec:variation}, we test the robustness of our results, and in Section~\ref{sec:compwithothers} compare our results with those of \citeal{loreig09} and others.  Lastly, in Section~\ref{sec:postmerger}, we speculate on the further evolution of our remnants, considering in particular whether, as suggested by \citeal{vkercj10}, some might lead to type Ia supernovae.

\section{Code and Input Physics}
\label{sec:gasoline}

We simulate the mergers by placing non-rotating white dwarfs in a circular orbit with an initial separation {\azero} chosen such that rapid mass transfer begins immediately.  We then follow the merger for six orbits, at which time the remnant has become approximately axisymmetric.  As in prior work, the morphology of all merger remnants is similar, consisting of a dense, primarily degeneracy-supported center surrounded by a partly thermally-supported hot envelope (called a ``corona'' by \citeal{loreig09}) and a thick, sub-Keplerian disk.  We will use the terms ``core'', ``envelope'' and ``disk'' throughout this work.  We also quite often refer to both the core and envelope simultaneously as the ``core-envelope''.

We use simulation techniques and initial conditions that are standard in the field of WD merger simulations, both in order to compare with previous work, as well as to not introduce novel numerical effects into our simulations.  We detail our code and initial conditions below so that they can easily be reproduced.

\subsection{The SPH Code}
\label{ssec:sphcode}

With smoothed-particle hydrodynamics, one uses particles as a set of interpolation points to determine continuum values of the fluid and model its dynamics.  SPH is a Lagrangian method, meaning movement is automatically tracked, and regions of high density contain more particles and therefore are automatically more resolved.  Moreover, SPH inherently conserves angular momentum in three dimensions, which is difficult to reproduce in grid codes except under specific coordinate systems and symmetries.  SPH therefore allows one to efficiently simulate complex phenomena with a large range of lengthscales.  It has become the method of choice for merger simulations, and so we chose it as well.


For our simulations, we use Gasoline \citep{wadssq04}, a modular tree-based SPH code that was designed and has been used for a wide range of astrophysical scenarios, from galaxy interactions to planet formation.  It aims for tight controls on force accuracy and integration errors.  Gasoline implements the \cite{hernk89} kernel -- we use 100 neighbors -- and uses the asymmetric energy formulation (\citeauthor{wadssq04}, Eqn. 8) to evolve particle internal energy.  In our simulations, total energy is on average conserved to 0.3\%, and angular momentum to 0.006\%.

By default, Gasoline uses the usual Monaghan and Gingold formulation for artificial viscosity (see \citealt{mona05}), together with a Balsara switch (a standard feature of WD merger SPH simulations) to reduce viscosity in non-shocking, shearing flows.  \cite{guerig04} found that such a prescription did not reduce viscosity sufficiently, resulting in excess spin-up of the remnant core and associated shear heating.  \cite{yoonpr07}, in addition to a Balsara switch, used variable coefficients for the linear and quadratic viscosity terms in the SPH equations of motion and energy, setting these values to $\alpha\,=0.05$ and $\beta\,=0.1$, respectively, where shocks are absent, and around unity where they are present.  A similar formulation was used in \cite{dan+11,dan+12}.  Since Gasoline includes it as well, we have used it for our study.  Excess viscosity nevertheless remains a potential problem;  we investigate its effects further in Sec.~\ref{ssec:viscprescrip}.

We modified Gasoline to include support for degenerate gas through the Helmholtz equation of state (EOS)\footnote{Available at \url{http://cococubed.asu.edu/} .} \citep{timms00}.  This code, also used in \cite{rask+12} and \cite{dan+12}'s simulations, interpolates the Helmholtz free energy of the electron-positron plasma, along with analytical expressions for ions and photons, to determine pressure, energy and other properties from density and temperature.  It is fast, spans a large range of density and temperature, and has, by construction, perfect thermodynamic consistency.  To obtain quantities as a function of density and internal energy, we utilized a Newton-Raphson inverter.  To keep the energy-temperature relation positive-definite, we did not disable Coulomb corrections in cases where total entropy became negative.


Gasoline keeps track of the internal energy of particles, using it to determine other thermodynamic properties for fluid evolution.  A particle's energy will naturally fluctuate due to noise, but for nearly zero-temperature particles this could result in their energy dipping below the Fermi energy.  In such situations we keep the pressure at the Fermi pressure, while letting the energy freely evolve.  A consequence of the floor is that a small amount of excess energy is injected into the system through mechanical work, which eventually manifests as additional thermal energy.  The accumulated energy over a simulation is typically a small fraction of the internal energy, and therefore does not significantly affect the dynamics of the merger or most properties of the remnant.  In cold, degeneracy-dominated material, however, a small change in internal energy corresponds to a large temperature change, at times comparable to the physically expected values, and thus the temperatures near the centers of some of our simulations have been affected.  We characterize this spurious heating in Sec.~\ref{ssec:spheat} and show that it does not unduly affect our work's conclusions.  However, it makes it difficult to run much longer simulations.

We also place an energy floor at half the Fermi energy.  This is to prevent particle energies from approaching zero (and consequently calling for tiny timesteps), which under rare circumstances occurs when particles perform a great deal of mechanical work.  We find this happens primarily for particles that are flung out of the system by the merger and are cooling rapidly, and therefore are confident it has only a very minor effect on our simulations.

In our work, we ignore outer hydrogen and helium layers, composition gradients, and any nuclear reactions.  This is mainly because previous work has found that nuclear processing was unimportant during the merger.  For instance, \citetalias{loreig09} found fusion released $\sim\!10^{41}$ erg for their 0.6 - 0.8 {\Msun} merger, orders of magnitude smaller than the $\sim\!10^{50}\,$erg binding energy of the remnant.  Only for mergers involving very massive, $\gtrsim\!0.9\,M_\odot$ WDs might this assumption break down, with the possibility of carbon detonations arising (\citealt{pakm+10,pakm+11,pakm+12}; but see \citealt{rask+12,dan+12}).  Similarly, \cite{rask+12}, who included standard helium envelopes of $\sim\!1-2$\% of the WD mass in their simulations, found that only for accretors with masses above $\sim\!1\,\Msun$ did it make a substantial difference: a helium detonation would inject $\sim\!10^{49}\,$erg into the merger remnant.  While this led to additional heating, it was insufficient to trigger much carbon burning or unbind any portion of the remnant (helium detonations have also been found for lower-mass accretors with CO-He hybrid donors; \citealt{dan+12}).

\subsection{Initial Conditions}
\label{ssec:initcond}

We created spherical white dwarfs using pre-relaxed cells of particles rescaled to follow the appropriate enclosed mass-radius relation determined using the Helmholtz equation of state.  We assumed a composition of 50\% carbon and 50\% oxygen by mass, and a uniform temperature of~$5\times10^6\,$K.  The stars were then relaxed in Gasoline for 81 s ($\sim$10 - 40 dynamical times, depending on the white dwarf mass) with thermal energy and motion damped (to~$5\times10^6\,$K and 0 cm s$^{-1}$, respectively) during the first 41 s, and left free during the remaining 40 s.  Particle energy noise prevented cooling of $\gtrsim\,5\times10^6\,\gcc$ material to below $10^7$ K.  We checked that the density profile of each star after relaxation was consistent with the solution from hydrostatic equilibrium, and found this was the case -- central densities, for example, agreed to within 2\%.  The radii of the relaxed stars, as defined by the outermost particle of a relaxed WD, on the other hand were on average about 7\% too small, reflecting our inability to model the tenuous WD outer layers\footnote{Our relaxed WDs also show evidence of sub-kernel radial banding of particles, which does not appear in any interpolated quantities.  We do not believe this banding has an effect on our simulations except for a possible reduction in effective resolution, but will investigate remedies in future work.}.

We used a constant particle mass of $10^{28}\,$g, so that a $0.4\,\Msun$ WD has $8\times10^4$ particles, and a $1.0\,\Msun$ WD has $2\times10^5$.  These numbers are similar to those used by \citetalias{loreig09} and \cite{yoonpr07}, and exceed the $\sim2\times10^4$ particles per star used by \cite{dan+12}. \cite{rask+12} performed a resolution test for a merger of two $0.81\,\Msun$ WDs, varying the number of particles per star from $10^5$ to $2\times10^6$.  They found differences of $\sim\!2\%$ in the mass of the core plus envelope, disk half-mass radius, and inner disk rotation frequency.  The one qualitative difference they found was that at their highest particle resolution, the WDs failed to break symmetry and disrupt (note that they assumed co-rotating WDs, making such a stable contact configuration possible).  We perform our own test in Sec.~\ref{ssec:restest} and find similar results.

We relaxed 0.4, 0.5, 0.55, 0.6, 0.65, 0.7, 0.8, 0.9 and 1.0 {\Msun} white dwarfs, and combined them in all possible permutations to form our parameter space of binaries.  These values were chosen to represent the range of possible CO WD masses, with greater resolution near the empirical peak at $\sim\!0.65\,\Msun$ of the mass distribution of (single) CO WDs \citep{tremb09}.  We also performed additional simulations with 0.575 - 0.65, 0.625 - 0.65 and 0.64 - 0.65$\,\Msun$ binaries to explore the outcomes of similar-mass mergers.  We thus simulated 48 mergers in total.

We placed two relaxed, irrotational WDs in a circular orbit.  We chose the initial separation {\azero} such that the donor WD just fills its Roche lobe, taking the location of the donor's outermost particle as its radius and using the Roche lobe approximation (for a synchronized binary) from \citet{eggl83}.  

This simple initial condition is similar to that of \cite{pakm+10}, and implies that the binary system as a whole is not equilibrated.  Therefore, as the simulation begins, the two WDs react to the tides, become stretched, and strong Roche lobe overflow ensues because the donor overshoots its Roche radius (in a widely separated binary, the donor would start to pulsate).  As a result, the donor disrupts after just one to two orbits.  For synchronized binaries, \cite{dan+11} showed that the onset of mass transfer is much more gentle if the WDs are relaxed in the binary potential, disruption occurring only after several dozen orbital periods.  They also showed that this results in systematic changes in the merger remnants.  It is not clear whether the same will hold for unsynchronized binaries, since the accretion stream hits a surface that, in its frame, counterrotates, and therefore accretion is always much less gentle than for synchronized WDs.  The difference is particularly dramatic for similar-mass binaries, where, in the synchronized case, the WDs can come into gentle contact, while in the unsynchronized case, any contact is violent.  Unfortunately, it is difficult to test the effect of proper equilibration for unsynchronized binaries, since one has to relax to non-trivial initial conditions.  A better approximation was attempted by \citeal{loreig09} and \cite{guerig04}, who started their WDs further out and reduced the separation artificially until mass transfer began.  In their simulations, disruption still followed very quickly.  Given that, and wanting to avoid any partial synchronization, we kept our simpler setup, and tested it by running simulations with varying~{\azero}.  We will discuss these tests in Sec.~\ref{ssec:varyingazero} and compare our results with those of others in Sec.~\ref{sec:compwithothers}.

\subsection{Merger Completion Time}
\label{ssec:mergercomplete}

It is difficult to decide when a merger is ``complete'', since for some cases remnant properties continue to evolve long after the two WDs coalesce, with (artificial) viscosity redistributing angular momentum and heating the remnant.  As a visually inspired criterion, we decided initially to use the degree of non-axisymmetry, continuing simulations until they were less than 2.5\% non-axisymmetric, as measured from the ratio of zeroth to largest non-zero Fourier coefficient of particles binned in azimuth.  However, this had its own issues: in dissimilar-mass mergers --  where most of the particles are in the accretor, already roughly axisymmetric following the merger -- our convergence criterion was achieved while the outer disk was still obviously non-axisymmetric.  In equal-mass mergers, which are inherently more axisymmetric, completion also was too soon, before the densest material had reached the center of the remnant.

For the majority of our systems, however, the time required to reach 2.5\% non-axisymmetry was roughly constant in units of the initial orbital period, at $6.1\pm1.2$.  For about the same time, axisymmetry was also achieved (by subjective visual inspection) for both dissimilar-mass mergers (except, in extreme dissimilar-mass cases, the outermost regions of their disks) and for equal-mass mergers (where the densest material had reached the center).  We therefore use 6 orbital periods of the initial binary as the completion time of our simulations.  In Sec.~\ref{ssec:runninglonger}, we discuss the effect of continuing our simulations for 2 further orbital periods.

\section{Results}
\label{sec:results}

With our 48 simulated mergers in hand, we try to determine scaling relations of global quantities such as the remnant and disk mass, highest temperature, etc., and look for homologies in the remnant profiles.  For our analysis, we use a cylindrical $(\varpi,\phi,z)$ coordinate system centered on the remnant core.  Properties on the equatorial $(\varpi,\phi)$ plane -- defined as the original orbital plane -- are averaged over $\phi$ using particles within $\frac{1}{2}\hz$ of the equatorial plane, where \hz\ is the remnant's rotational axis ($\varpi = 0$) central scaleheight (see Sec.~\ref{sssec:structuraltrends}).  Properties along the rotational $(z)$ axis are averaged within a cylinder $\varpi<\frac{1}{2}\hz$.  We use $\frac{1}{10}\hz$ as the bin size along both the equatorial plane and rotational axis.  We determine properties mostly as a function of enclosed mass $M(r)$, which we define spherically\footnote{Arguably, enclosed mass is more properly defined within equipotential surfaces, but this makes comparison with other simulations harder.  For dissimilar-mass mergers, the difference is slight.}.  Thus, we show, e.g., equatorial plane temperature $T(\varpi)$ as a function of $M(r=\varpi)$, the mass enclosed within a sphere with radius $r=\varpi$.

\subsection{Representative Mergers}
\label{ssec:samplingofmergers}

\begin{figure*}
\centering
\includegraphics[width=2.0\columnwidth]{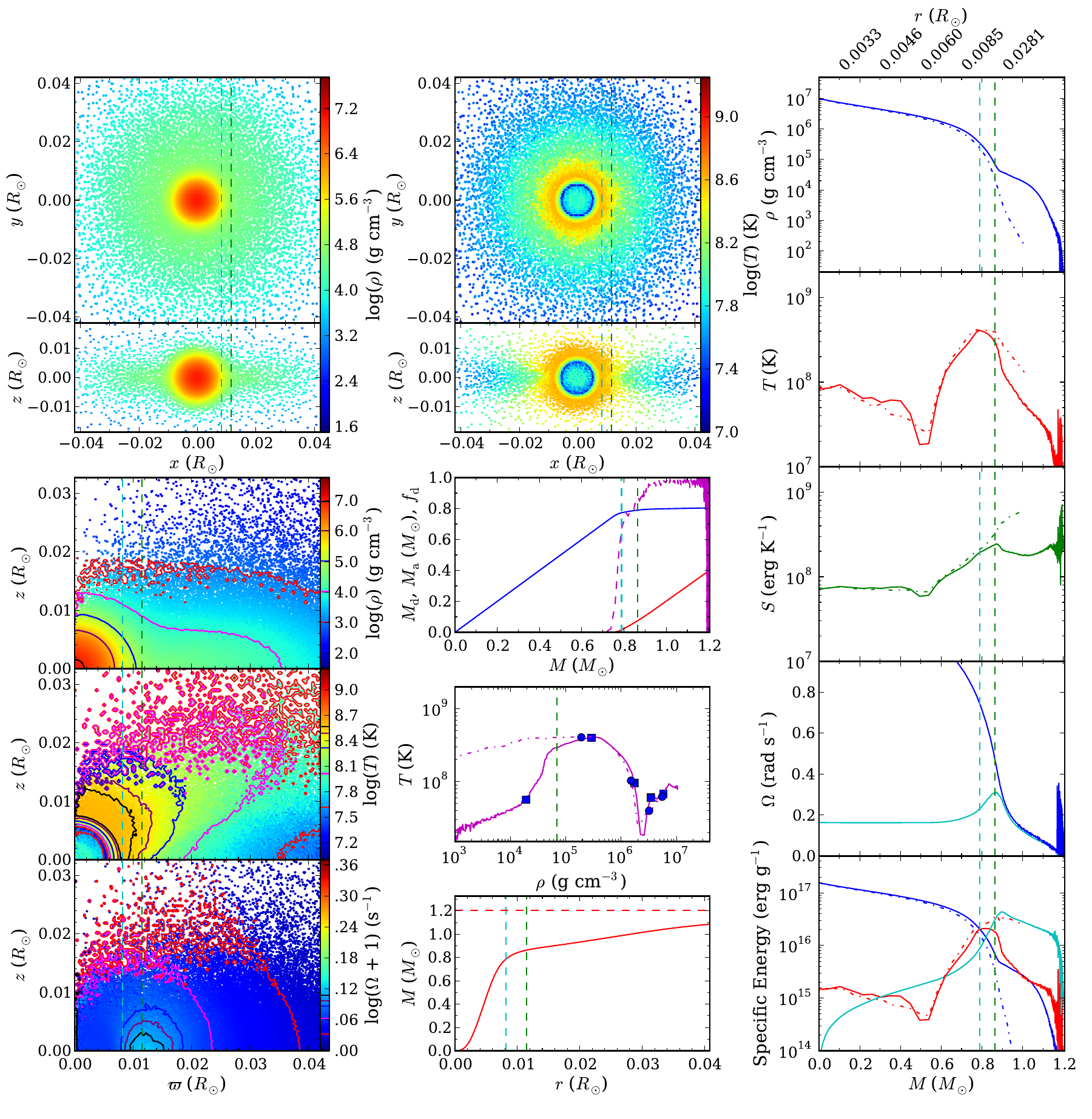}
\caption{Structure of a 0.4 - 0.8 {\Msun} merger remnant, representing the general outcome of a merger of white dwarfs with dissimilar mass.  Upper left and middle -- binned maps of density $\rho$ and temperature $T$ along slices in the $xy$ and $xz$-planes.  Lower left -- binned maps and contours of density, temperature, and angular frequency $\Omega$ in the $(\varpi,z)$ plane, averaged over cylindrical coordinate $\phi$ and over $\pm z$ (with 1 added to $\Omega$ to avoid problems with the logarithmic intensity scale).  Middle -- enclosed masses of donor and accretor material {\Md} and {\Ma} (solid red and blue, resp.), and fraction of donor material \fdon\ at a particular mass shell (dashed magenta).  Middle, one but lowest -- temperature-density profile with enclosed masses in 0.2\,\Msun\ increments indicated, both along the equatorial plane (solid curve, squares) and along the rotational axis (dot-dashed curve, circles).  Middle, bottom -- enclosed mass as a function of $r$, with the total mass indicated by the horizontal dashed red line.  Right-hand column, top to bottom - density, temperature, entropy, angular (cyan) and Keplerian (blue) frequency, and degeneracy (blue), thermal (red) and rotational (cyan) specific energies as a function of enclosed mass $M$, both along the equatorial plane and along the rotational axis (solid and dot-dashed curves, respectively).  In all graphs, the start of the disk (where the centrifugal acceleration equals half the gravitational one) and the equatorial radius (or mass enclosed within) of maximum temperature are marked by vertical green and blue dashed lines, respectively. [{\em See the electronic edition of the Journal for Figs.~1.1--1.48}.]}
\label{fig:mergersampling1}
\end{figure*}

\begin{figure*}
\centering
\includegraphics[width=2.0\columnwidth]{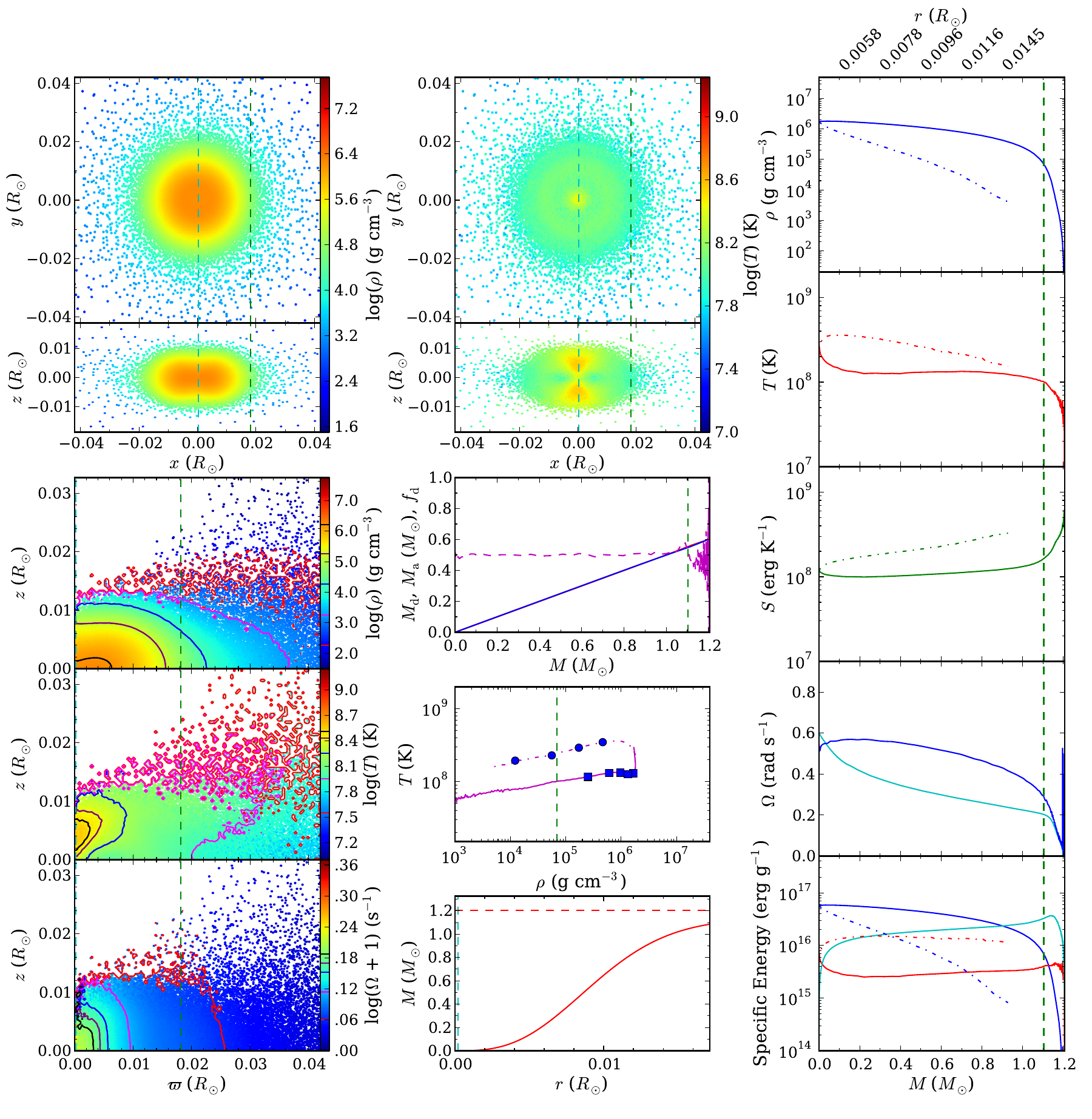}
\caption{As Fig. Set~\ref{fig:mergersampling1}, but for a 0.6 - 0.6 {\Msun} merger remnant, representing the general outcome of a similar-mass merger.}
\label{fig:mergersampling2}
\end{figure*}

As found for previous simulations, qualitatively the most important factor controlling the merger outcome is whether the WD masses are ``dissimilar'' or `` similar''.  In the former case, where the donor is significantly less massive than the accretor, only the donor overflows its Roche lobe,\footnote{The lower mass WD is larger and thus always fills its (smaller) Roche lobe first.} is disrupted, and accretes onto the accretor.  The accreted material is heated on impact, lifting degeneracy.  Hence, the merger remnant consists of a partly non-degenerate hot envelope and small, thick sub-Keplerian disk, both surrounding a cold core containing the largely unaffected accretor.

In the latter case of a similar-mass merger, there is a large degree of mixing between the two stars.  For exactly equal masses, both stars are disrupted simultaneously, and their accretion streams impact each other near the system's barycenter.  Material from the centers of both stars initially forms a thick, cold, dense torus orbiting the barycenter; this torus slowly shrinks due to viscous drag, pushing the accretion stream material above and below the equatorial plane.  When the stars have slightly different mass, the lower-mass one disrupts first, forming an accretion stream (or series of streams) that mixes with accretor material down to the center of the accretor (regardless of whether or not the other also disrupts).

We show the differences between similar and dissimilar-mass merger remnants using two representative examples in Figs.~\ref{fig:mergersampling1} and~\ref{fig:mergersampling2}: a 0.4 - 0.8 {\Msun} highly dissimilar and a 0.6 - 0.6 {\Msun} equal-mass merger, respectively.  One sees that the remnant morphologies are very different, consistent with previous work.  The 0.4 - 0.8 {\Msun} merger features a cold, nearly non-rotating and thus spherically symmetric remnant core, surrounded by a hot envelope with roughly equal degeneracy and thermal support, which itself is surrounded on the equatorial plane by a rotationally supported non-degenerate thick disk that holds most of the angular momentum.  The accretor forms the core, largely undisturbed by the merger, while the envelope and disk are composed almost entirely out of donor material.  The hottest points are on the interface between the core and the envelope.\footnote{The higher temperatures near the core are spurious; see Sec.~\ref{ssec:spheat}}  The 0.6 - 0.6 {\Msun} remnant, on the other hand, has a massive, hot, partly rotationally supported and thus ellipsoidal core, and a very small but thick disk, both of which consist of material from both stars.  No distinct envelope is formed.  The hottest points are within the remnant core, just above and below the equatorial plane, arising from accretion stream material pushed out by the shrinking dense torus.  \notetoeditor{The entire figure set can be found at http://astro.utoronto.ca/$\sim$cczhu/FigSet.tar.gz} 



\begin{figure}
\centering
\includegraphics[angle=0,width=1.0\columnwidth]{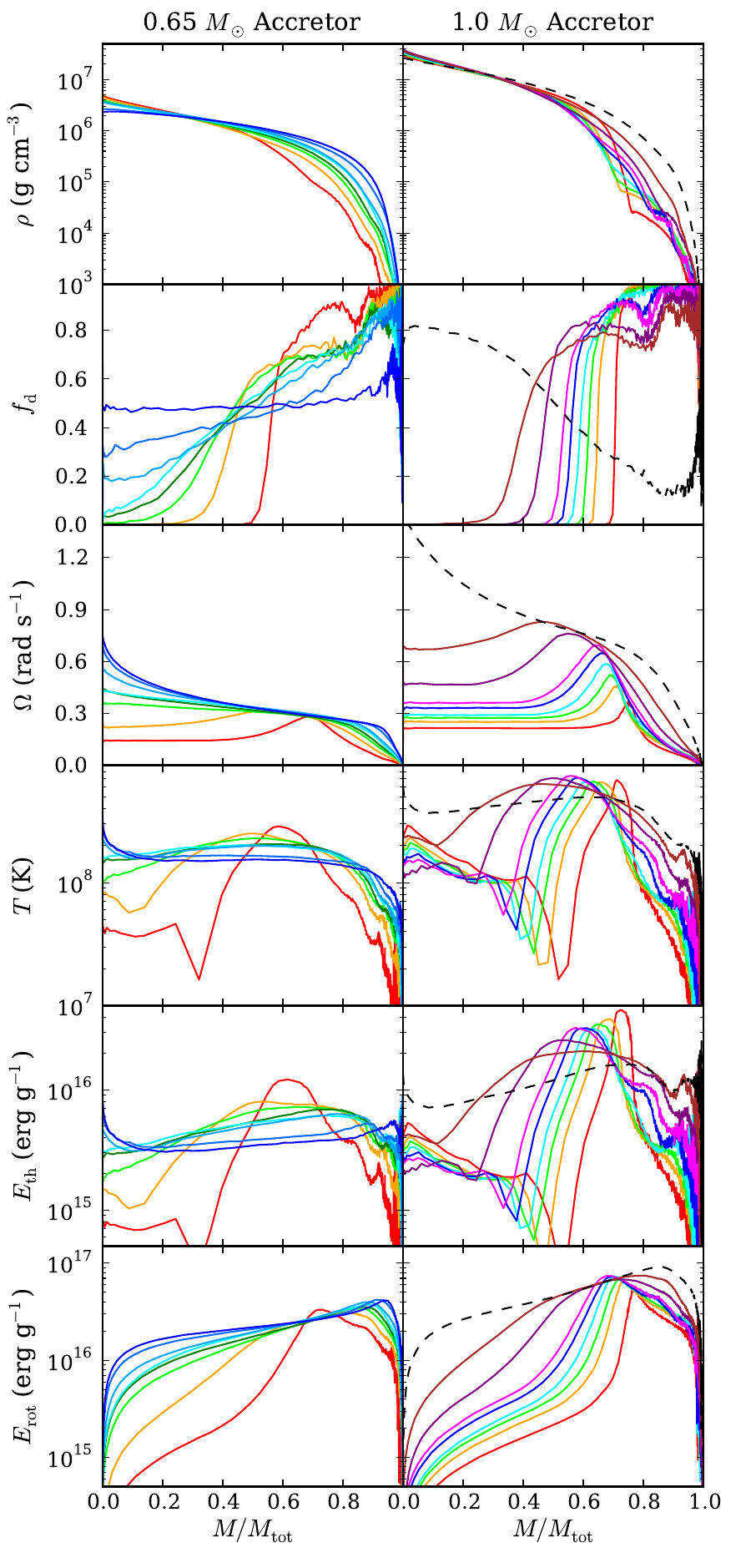}
\caption{Properties of mergers with 0.65\,\Msun\ (left) and 1.0\,\Msun\ (right) accretors, for donor masses of 0.4 (red), 0.5 (orange), 0.55 (lime), 0.575 (green), 0.6 (cyan), 0.625 (light blue), 0.64 (blue), 0.65 (dark blue), 0.7 (magenta), 0.8 (purple), 0.9 (brown), and 1.0\,\Msun\ (black).  Shown are, from top to bottom, density $\rho$, fraction of donor material \fdon, angular frequency $\Omega$, temperature $T$, specific thermal energy $E_{\rm th}$, and specific rotational energy $E_{\rm rot}$, all as a function of fractional enclosed mass $M/M_{\rm tot}$.  All properties are determined along the equatorial plane, except for \fdon\ which is defined spherically.  The 1.0 - 1.0 {\Msun} merger (dashed black line) is an outlier; see text.}
\label{fig:constacc}
\end{figure}

A good way to visualize how mergers transition between dissimilar and similar-mass is to look at changes in the remnant properties with varying donor mass.  In Fig. \ref{fig:constacc}, we show curves for accretors of 0.65 (left) and 1.0\,\Msun (right).  One sees that remnants of highly dissimilar-mass mergers, with mass ratio $\qm\equiv M_\mrm{d}/M_\mrm{a} \lesssim0.5$, have properties resembling the 0.4 - 0.8 {\Msun} merger: their donor and accretor barely mixed, their temperature curves have off-center hot plateaus, and their angular velocity profiles feature an off-center bump.  The equal-mass, $\qm=1$ cases resemble the 0.6 - 0.6 {\Msun} remnant: they have flat temperature profiles and centrally peaked angular velocity profiles.  Intermediate cases have intermediate profiles, with the bumps in the temperature and angular velocity profiles widening with increasing {\qm}.  The 0.4 - 0.8 {\Msun} and 0.6 - 0.6 {\Msun} remnants therefore lie at the extremes of what merger remnants look like.

\begin{figure*}
\centering
\includegraphics[angle=0,width=2.0\columnwidth]{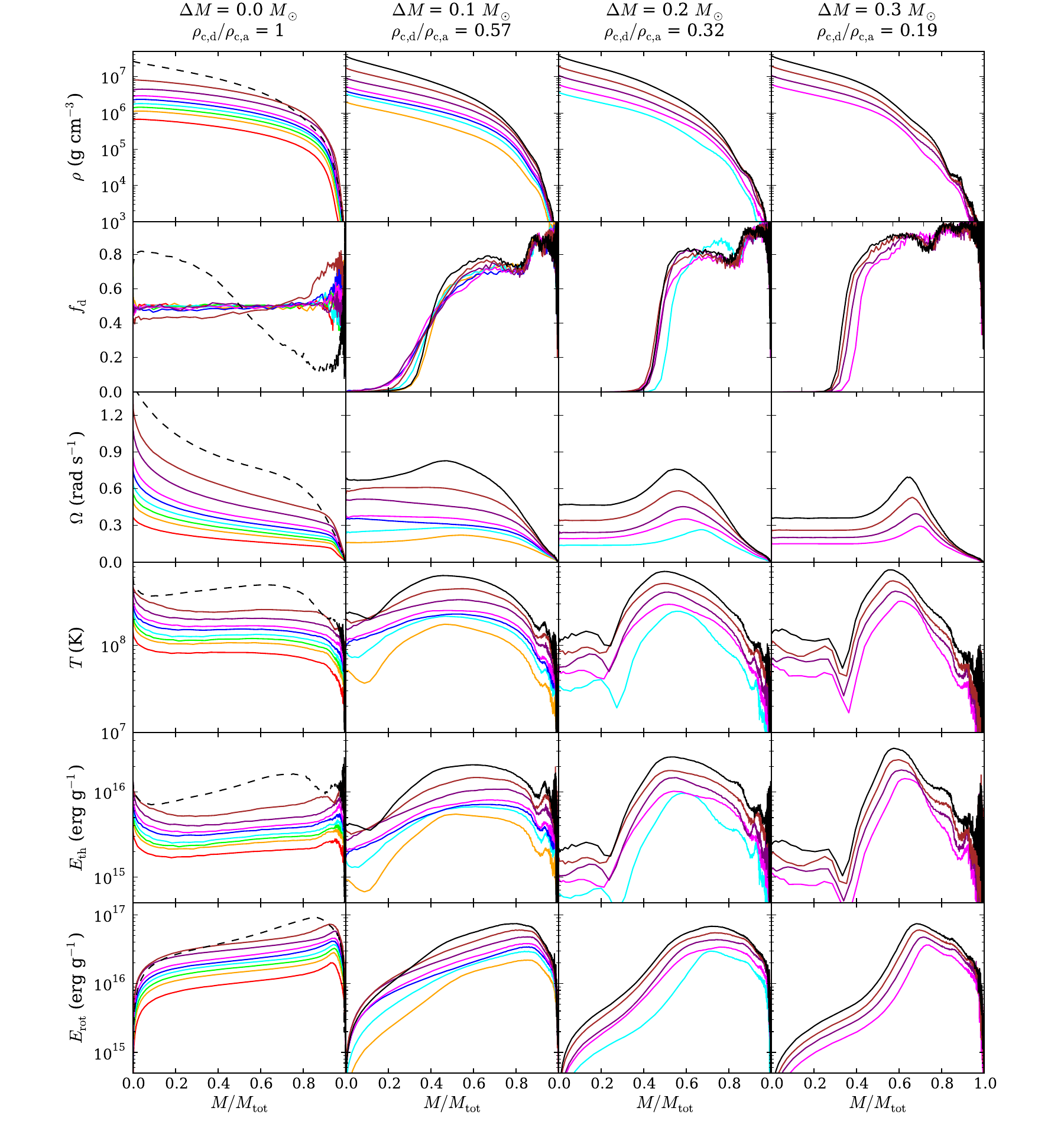}
\caption{Dependence of the properties of mergers on mass difference, with, from left to right, $\Delta M \equiv $ {\Ma} - {\Md} = 0.0, 0.1, 0.2, and $0.3\,\Msun$ mergers.  Properties shown, coloring, and line styles are as in Fig.~\ref{fig:constacc}, except color represents accretor mass.}
\label{fig:deltamcomp}
\end{figure*}

The similarity between some of the curves for the 0.65 and 1.0\,\Msun\ accretors in Fig. \ref{fig:constacc} suggests a homology.  The similarity is closest for mergers with the same mass difference $\Delta M$, as can be seen in Fig. \ref{fig:deltamcomp}.
For equal-mass mergers, all profiles are similar, simply scaled by a factor that depends on the total mass (except the 1.0 - 1.0 {\Msun} merger; see below).  As $\Delta M$ increases, the profiles are slightly less similar: with increasing total binary mass, the degree of mixing decreases, and the temperature and angular velocity maxima drift to slightly lower fractional enclosed mass.  Nevertheless, the profiles still resemble one another far more closely than they resemble curves with other $\Delta M$.  The same holds for profiles along the rotational axis.

It may seem surprising that the controlling parameter between these approximate homologies is the mass difference $\Delta M$ rather than the mass ratio {\qm}.  Empirically, however, the case is clear: e.g., the 0.4 - 0.5 (second column, yellow) and 0.8 - 1.0 {\Msun} (third column, black) mergers have the same {\qm}, but different $\Delta M$, and their structures clearly differ from one another.  The same is true for the 0.4 - 0.6 (third column, cyan) and 0.6 - 0.9 (fourth column, brown) {\Msun} mergers.  As we discuss below, the similarity of mergers of similar $\Delta M$ likely reflects the close relation between the ratio of central densities and mass difference.  

Before discussing the homologies and trends further, we should note the one dramatic exception.  The 1.0 - 1.0 {\Msun} simulation differs fundamentally from its fellow $\Delta M\,=\,0$ mergers.  During the evolution of this system, unlike for all other equal-mass mergers, one WD was fully disrupted before the other, and as a result material from one star (arbitrarily designated the ``donor'' before the start of simulation, hence the ``inverted'' mixing profile in Figs.~\ref{fig:constacc} and~\ref{fig:deltamcomp}) preferentially resides near the center of the remnant.  This system also often appears as an outlier in Sec.~\ref{ssec:mergertrends} below.  The 0.9 - 0.9 {\Msun} merger also did not have equal mixing between the two stars, though the difference is much smaller.  \cite{rask+12} noticed the same effect in their simulations, and concluded it reflected the fact that more massive WDs are much more concentrated and therefore harder to disrupt.  This seems a likely explanation.


\subsection{Merger Trends}
\label{ssec:mergertrends}

A major goal of our work is to establish how various global properties of the merger remnant, such as remnant core and disk mass, maximum temperature, and maximum angular velocity, vary as a function of accretor and donor mass.  By quantifying these trends, we hope to help develop a parametrized model of merger remnants.  Before discussing trends, however, we stress that they are necessarily \textit{approximate} - second order effects, numerical noise and our choice of stopping time all affect the remnant properties.  Moreover, while integrated values like total thermal energy do not fluctuate from timestep to timestep, values at specific points in the remnant do (as noted the following sections).  For instance, the mass enclosed within the radius of peak equatorial temperature becomes ill-defined for similar-mass mergers because these have rather flat temperature profiles (Fig.~\ref{fig:constacc}).  To partly mitigate these fluctuations, the values presented below were determined by averaging frames from the simulation over an eight second span, centered on the time corresponding to six orbits of the initial binary.

\begin{figure}
\centering
\includegraphics[angle=0,width=1.0\columnwidth]{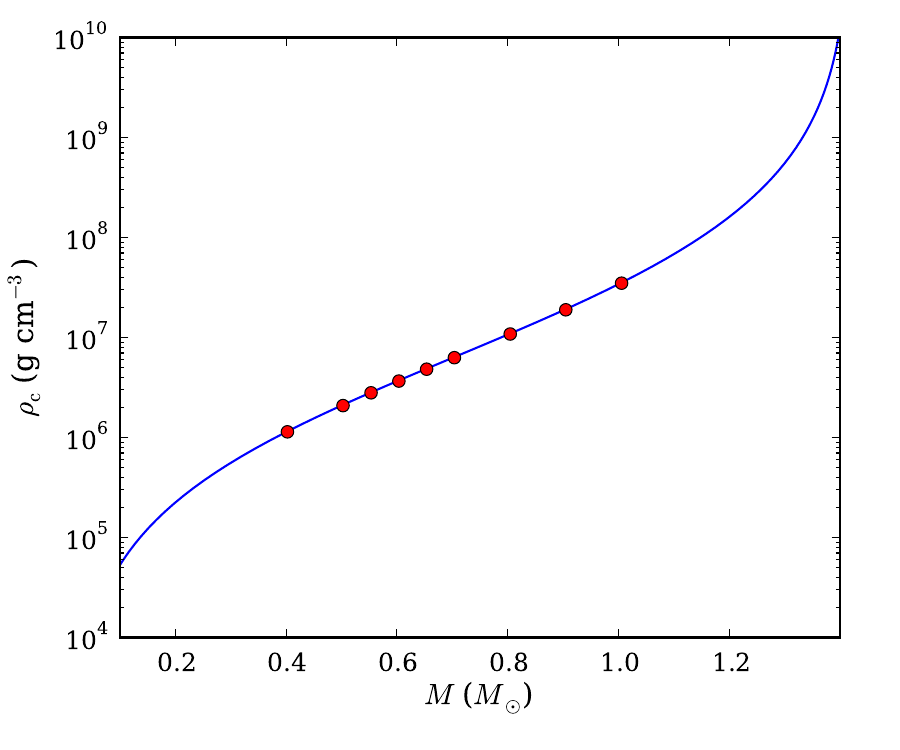}
\caption{Relation between central density \rhoc\ and mass $M$ for carbon-oxygen white dwarfs, showing both the results of relaxing white dwarf models in Gasoline (red points), and integrating hydrostatic equilibrium directly for spherically symmetric, non-rotating CO WDs with $T=5\times 10^6\,$K (blue line).  For the mass range considered, the central density depends roughly exponentially on mass.}
\label{fig:mrho}
\end{figure}

As might be expected from the approximate homologies described above, we found that many properties scaled well with $\Delta M$.  Of course, a scaling with a dimensional mass difference makes little sense; we believe its success reflects the fact that over the range of 0.4 -- 1.0\,{\Msun}, the central density {\rhoc} depends approximately exponentially on mass, with $\rho_{c} \simeq 3.3\times10^7{\rm\,g\,cm^{-3}} \exp[5.64(M/\Msun-1)]$ (see Fig.~\ref{fig:mrho}).  Hence, a given mass difference $\Delta M$ corresponds to a given ratio of central densities, \rhocrat.  As argued in Sec.~\ref{sssec:masstrends}, \rhocrat\ has a straightforward interpretation: it characterizes the degree of mixing between the donor and accretor.  We therefore discuss trends as a function of $\qrho\equiv\rhocrat$ from hereon.  Where necessary, we refer to the mass ratio as \qm.


\subsubsection{What Constitutes Similar-Mass?}
\label{sssec:whatisequalmass}


As {\qrho} increases from a small value toward unity, the merger remnant's morphology shifts from resembling Fig.~\ref{fig:mergersampling1} (dissimilar-mass) to resembling Fig.~\ref{fig:mergersampling2} (equal-mass).  From Fig.~\ref{fig:constacc}, ones sees that there is no particular {\qrho} at which one transitions from ``dissimilar'' to ``similar.''  Nevertheless, we can determine a rough critical value of~{\qrho} that separates mergers in which the core is largely unaffected from those in which it is changed significantly, a separation that likely affects the outcome of post-merger evolution.

\begin{figure}
\centering
\includegraphics[width=1.0\columnwidth]{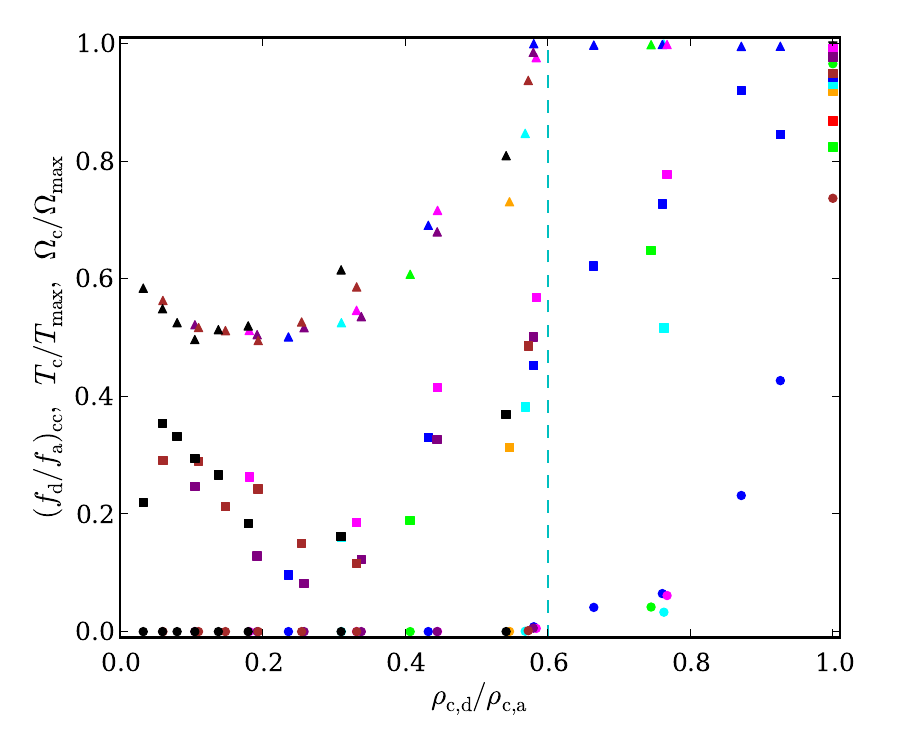}
\caption{Dependence of merger core properties on the ratio of the donor and accretor central densities, \rhocrat.  Shown are the ratio of central to maximum temperature $T_\mathrm{c}/T_\mathrm{max}$ (squares), central to maximum angular velocity $\Omega_\mathrm{c}/\Omega_\mathrm{max}$ (triangles), and central core donor to accretor mass fraction \fratio\ (circles), with colors representing different accretor masses, encoded as in Fig.~\ref{fig:constacc}.  The vertical line marks $\qrho\equiv\rhocrat=0.6$, where $\Omega_\mathrm{c}/\Omega_\mathrm{max}$ reaches unity, \fratio\ becomes non-zero, and $T_\mathrm{c}/T_\mathrm{max}\simeq0.5$.  We suggest it separates ``dissimilar'' from ''similar'' mass mergers.}
\label{fig:eqmass}
\end{figure}

To determine the critical value, we show in Fig.~\ref{fig:eqmass} the ratio of central to maximum temperature, $T_\mrm{c}/T_{\rm max}$, central to maximum angular velocity, $\Omega_\mrm{c}/\Omega_{\rm max}$, and the fraction of donor to accretor material within the central core, \fratio, where we define the central core as a sphere with radius \hz.  All three properties are measures of the extent to which the core has been affected: mixed regions tend to be hotter and more spun up, and contain material from both stars.

From Fig.~\ref{fig:eqmass}, one sees that $\Omega_\mrm{c}/\Omega_\mathrm{max}$ approaches unity at $q_\rho\simeq0.6$; at higher values, the angular velocity profile has a plateau or central peak rather than an off-center bump.  Also at $q_\rho\simeq0.6$, \fratio\ starts to deviate from zero, i.e., donor material begins to penetrate the central core.  The temperature points show the transition is not abrupt: $T_\mathrm{c}/T_\mathrm{max}$ starts to deviate from its downward trend (which reflects spurious heating in the most dissimilar-mass mergers; Sec.~\ref{ssec:spheat}) around $q_\rho\simeq0.3$ and continues to increase until $q_\rho=1.0$; at $q_\rho\simeq0.6$, $T_\mathrm{c}/T_\mathrm{max}\simeq0.5$.  Overall, this suggests that while the dependence is gradual, the morphology changes most around $\qrho\simeq0.6$.  This conclusion is confirmed by looking at the two-dimensional remnant temperature structures (Figs.~\ref{fig:mergersampling1} and~\ref{fig:mergersampling2}).  At $q_\rho\ll0.6$, the remnant core has a large, spherically symmetric cold region, the nearly unperturbed accretor.  This cold region shrinks with increasing \qrho, and at $q_\rho\simeq0.6$, spherical symmetry is broken.  For still larger \qrho, the cold region becomes a flat slice sandwiched between hotspots off the equatorial plane.

Given the above, we define ``similar-mass'' mergers as those with donor to accretor central density ratio $q_\rho>0.6$, and ``dissimilar-mass'' mergers as those with $q_\rho<0.6$.  This critical density ratio corresponds to a mass difference $\Delta M\simeq0.1\,\Msun$.

\subsubsection{Structural Trends}
\label{sssec:structuraltrends}

Here and in the following subsections of \ref{ssec:mergertrends}, we describe various trends of remnant properties in detail, hoping to help attempts to interpolate between different simulations and motivate analytical and semi-analytical depictions of the merger.  Readers not requiring this level of detail may wish to skip to Sec. \ref{ssec:qualitative}.  We begin our discussion of trends with size and density parameters.

\begin{figure}
\centering
\includegraphics[angle=0,width=1.0\columnwidth]{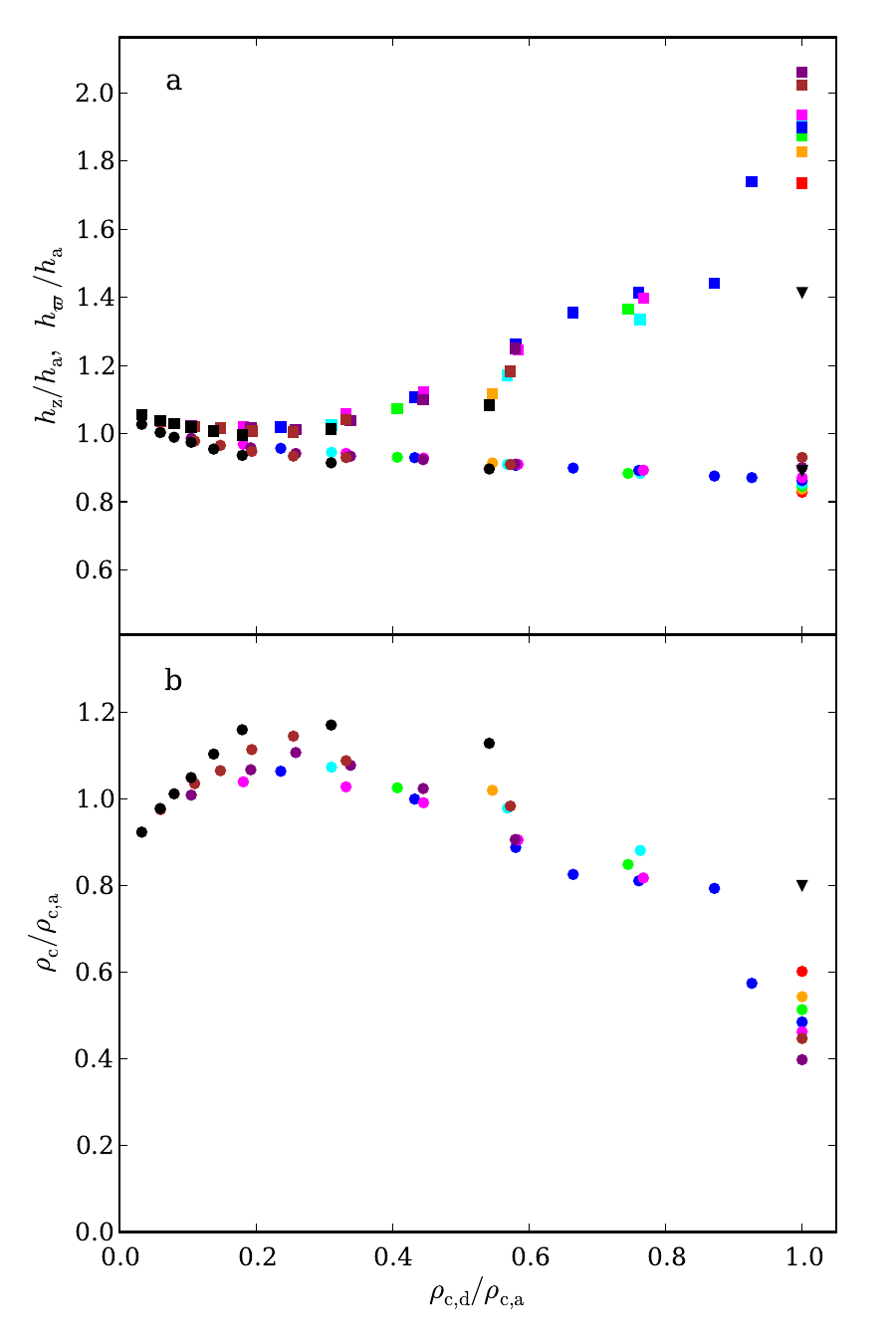}
\caption{Structural properties of mergers.  (a) Central scaleheights along the rotational axis (circles) and along the equatorial plane (squares) scaled to the scaleheight of the accretor, {\hz}/$h_\mathrm{a}$ and $\hxy$/$h_\mathrm{a}$.  (b) Central density of the merger remnant scaled to the central density of the accretor, {\rhocrhoct}.  Colors represent different accretor masses, encoded as in Fig.~\ref{fig:constacc}.  Triangles represent the outlying 1.0 - 1.0\,\Msun\ merger.}
\label{fig:structuretrends}
\end{figure}

\paragraph{The rotational axis central scaleheight.} We define the rotational axis central scaleheight \hz\ as the characteristic width $\sigma$ of a Gaussian fit to the density distribution along the $z$ axis at $\varpi = 0$.  \hz\ is a measure of the vertical extent of the remnant.  We find that the ratio $\hz/h_{\rm a}$, where $h_\mathrm{a}$ is the central scaleheight of the accretor, is reasonably well-approximated by,
\eqbegin
\frac{h_\mrm{z}}{h_\mathrm{a}} = 1.03 - 0.17\qrho^{1/2}
\qquad(\pm0.02),
\eqend
where the uncertainty listed in parentheses represents the root-mean-square (RMS) of the residuals around the approximation (see Fig.~\ref{fig:structuretrends}a).  For highly dissimilar-mass mergers, \hz\ approximately equals the scaleheight of the accretor, while for similar-mass mergers, \hz\ is lower due to rotational support.

The vertical scaleheight increases with increasing $\varpi$: the scaleheight at the location of maximum temperature, $h(T_{\rm max})$, ranges from \hz\ to 1.21\hz, and the scaleheight at maximum angular velocity $h(\Omega_{\rm max})$ ranges from \hz\ to 1.88\hz.  The prefactor for both heights increases with increasing accretor mass {\Ma} and decreasing \qrho. 

\paragraph{The equatorial plane central scaleheight.}  Similar to \hz\, we define \hxy\ -- the characteristic width of a Gaussian fit to the density distribution along the equatorial plane -- as a measure of the equatorial extent of the remnant.  The ratio $\hxy/h_{\rm a}$ can be parametrized by
\eqbegin
\frac{\hxy}{h_{\rm a}} = 0.96 + 0.89q_\rho^2
\qquad(\pm0.08),
\eqend
where we excluded the 1.0 - 1.0 {\Msun} merger remnant for our fit (see Fig.~\ref{fig:structuretrends}a).  The dependence on increasing \qrho\ reflects the increased rotational support of the remnant core.

\paragraph{The central density of the remnant.}  The central density, \rhoc, is always within a factor two of the central density of the accretor, \rhoca.  In Fig.~\ref{fig:structuretrends}b, one sees that for given accretor mass, \rhocrhoct\ increases with increasing \qrho\ for highly dissimilar-mass mergers due to increasing compression of the remnant core, but begins to decrease because of increasing rotational support around $\qrho\simeq0.3$.  We could not find a simple parametrization for these curves.  Note that for some systems, as we continue running our simulations \rhocrhoct\ continues to increase.  As discussed in Sec.~\ref{ssec:mergercomplete}, this is probably because artificial viscosity forces the merger remnant to undergo accelerated viscous evolution.



\subsubsection{Mass Distributions}
\label{sssec:masstrends}

The merger mixes material between the donor and accretor.  Here, we describe how this changes as a function of \qrho, as well as how the material is distributed between the pressure-supported core and envelope and the rotationally supported disk.

\begin{figure*}
\centering
\includegraphics[angle=0,width=2.0\columnwidth]{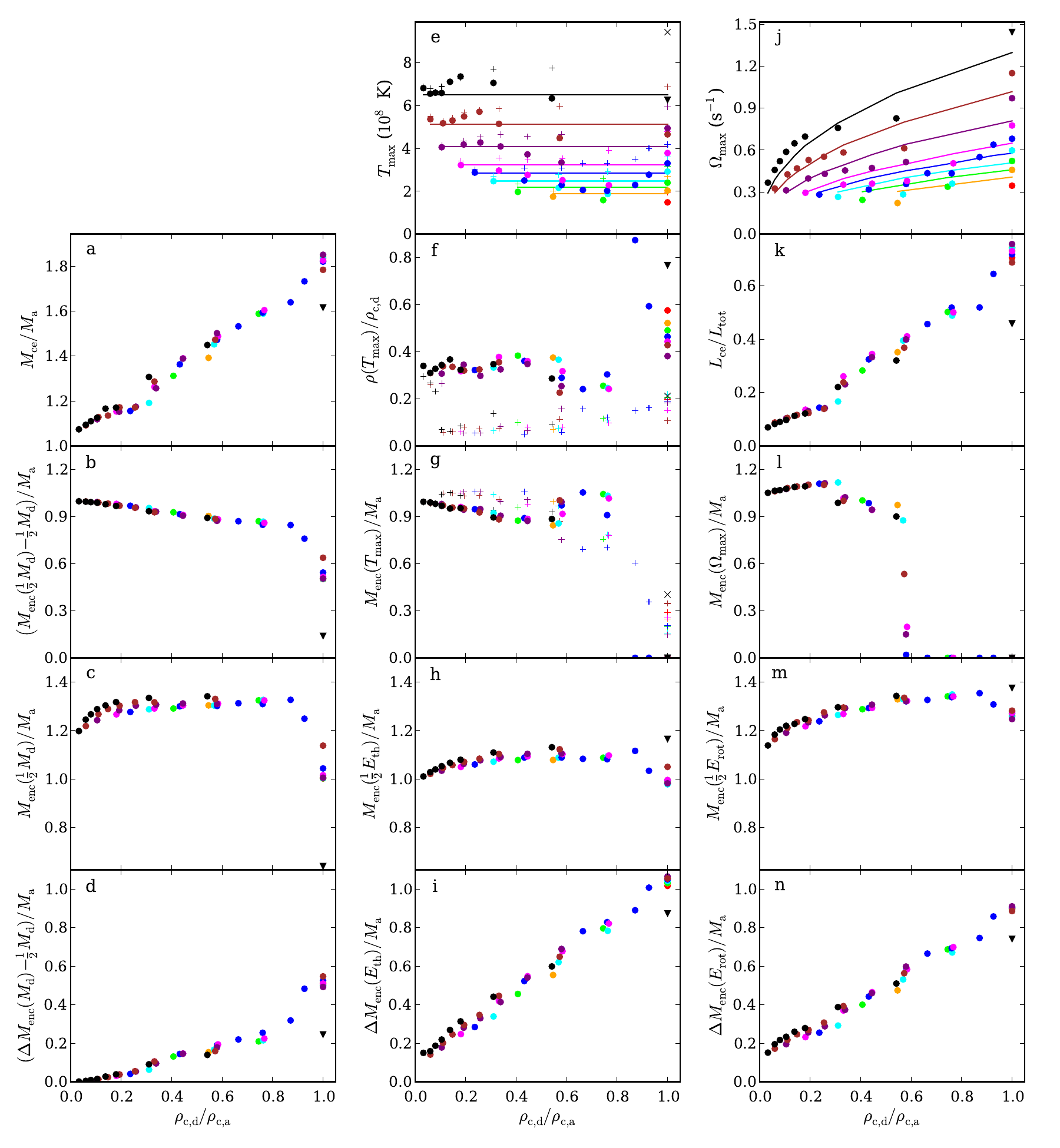}
\caption{Mixing, heating, and spin-up (left to right) for mergers. (a) Scaled mass of the remnant core-envelope (where scaling here and below is to the accretor mass).  (b) Fraction of the accretor within the mass enclosing half the donor mass.  (c) Scaled remnant mass enclosing half the donor mass.  (d) Fraction of the accretor mass within the region enclosing 25--75\% of the donor mass.  (e) Maximum equatorial temperature \Tmax\ (circles), with the approximation $\Tmax=0.20G\Ma m_\mathrm{p}/k_B\Ra$ overdrawn.  Maximum temperatures along the rotational axis are shown with crosses.  (f) Scaled density at the location of \Tmax\ (symbols as above).  (g) Scaled mass enclosed within the radius of \Tmax\ (symbols as above).  (h) Scaled mass enclosing half of the remnant thermal energy.  (i) Scaled mass of the region enclosing 25 -- 75\% of the remnant thermal energy.  (j) Maximum angular velocity \Omegamax\ (circles) with best fit $\Omegamax=3.8\Omega_\mrm{orb}$ overdrawn.  (k) Fraction of the angular momentum in the core-envelope.  (l) Scaled mass enclosed within the radius of maximum angular velocity.  (m) Scaled mass enclosing half of the total remnant rotational energy.  (n) Scaled mass of the region enclosing 25 -- 75\% of the remnant rotational energy.  Colors represent different accretor masses, encoded as in Fig.~\ref{fig:constacc}.  Triangles represent equatorial plane values, and x-marks rotational axis values, of the 1.0 - 1.0\,\Msun\ merger.}
\label{fig:restoftrends}
\end{figure*}

\paragraph{The masses of the core-envelope and disk.}  We formally define the core-envelope as the part of the remnant inside the inner disk radius $\varpi_\mrm{disk}$, i.e., that is supported primarily by pressure (degeneracy for the core, thermal for the envelope) and not rotation.  Since in every merger very little mass is ejected, a trend for either the core-envelope or the disk mass (\Mrem\ and \Mdisk, resp.) suffices to determine both.  The ratio of \Mrem\ to the accretor mass \Ma\ is well described by,
\eqbegin
\frac{\Mrem}{\Ma} = 1 + 0.81\qrho
\qquad(\pm0.03),
\eqend
if the 1.0 - 1.0 {\Msun} merger is neglected, and the fit's y-intercept is forced to unity.  See Fig.~\ref{fig:restoftrends}a.


\paragraph{The mass enclosing 50\% of the donor material.}  The further the donor penetrates, the smaller will be the mass enclosing half the donor's material, \MMfifty.  For mergers with $\qrho\lesssim0.8$, $\MMfifty/\Ma\,\simeq\,1.30$, with an RMS residual of 0.03 (Fig.~\ref{fig:restoftrends}c).  We present this trend mostly because we discuss similar thermodynamic and rotational enclosed masses, but it is somewhat difficult to interpret physically, since \MMfifty\ increases with donor mass but decreases with mixing, which also depends on donor mass.  The trend is easier to interpret using enclosed accretor mass rather enclosed total mass, as done below.

\paragraph{The accretor mass enclosing 50\% of the donor material.}  As a different measure of the depth to which the donor penetrates, we consider just the accretor material within the mass enclosing half the donor, $\MMfifty-\frac{1}{2}\Md$.  This should equal the accretor mass if the donor is deposited above the accretor, and half the accretor mass if the two stars are completely mixed.  For $\qrho\lesssim0.8$, it can be approximated by (Fig.~\ref{fig:restoftrends}b),
\eqbegin
\frac{\MMfifty-\frac{1}{2}\Md}{\Ma} = 1 - 0.190\qrho
\qquad(\pm0.009),
\eqend
where we forced the intercept to be unity.  In this regime, roughly half of the donor remains outside of the accretor, though the trend discussed next indicates that the other half which does penetrate the accretor is spread across a much larger region at higher \qrho.  When $q_\rho\gtrsim0.8$, the ratio drops sharply downward, indicative of the more thorough mixing expected for the similar-mass case.  However, the existence and exact location of this drop may be a function of initial conditions (see Sect.~\ref{ssec:varyingazero}).  

\paragraph{The region over which the donor is spread.} As a measure of the thickness of the region affected by the merger, we use the difference of the mass enclosing 75\% of the donor material with that enclosing 25\% of the donor material, i.e., $\MMthick = M_{\rm enc}(\frac{3}{4}\Md) - M_{\rm enc}(\frac{1}{4}\Md)$.  Since 50\% of the donor is within this range, $\MMthick - \frac{1}{2}\Md$ is a measure of the amount of accretor mixed with the donor.  For $q_\rho\lesssim0.8$, the ratio of the latter to the total accretor mass follows,
\eqbegin
\frac{\MMthick-\frac{1}{2}\Md}{\Ma} = 0.30\qrho
\qquad(\pm0.02),
\eqend
while for $q_\rho\gtrsim0.8$ the trend curls upward until it reaches 0.5, the value expected for completely mixed remnants.  See Fig. \ref{fig:restoftrends}d.

Combining the two above trends, we can formulate a qualitative picture of mixing.  For $q_\rho\lesssim0.8$, the donor can be thought of as being deposited onto the accretor and mixing with the accretor's outer layers, while for $q_\rho\gtrsim0.8$, the accretor also disrupts substantially, leading to a regime where both stars mix more uniformly.  The region over which the donor is spread, or thickness of the mixed layer, in both cases depends on {\qrho}, which suggests that the relative densities of the donor and accretor govern mixing, i.e., the donor mixes significantly with all accretor material up to some fraction of the central density of the donor.  Additional evidence of this will be seen in the thermodynamic trends below.  

One might consider an alternate picture in which the donor dredges up a constant fraction of its own mass in accretor material.  If this were the case, we would expect $(\MMthick-\frac{1}{2}\Md)/\Md$ to roughly be constant.  Our results, however, show that for $q_\rho\lesssim0.8$ this quantity is nearly a straight line that is close to zero for small \qrho\ ($(\MMthick-\frac{1}{2}\Md)/\Md = 0.35\qrho\pm0.02$).  This seems more consistent with mixing being determined by density.


\subsubsection{Energy Balance}

The energy balance of the remnants indicate their primary means of support.  Since the remnants are virialized, we consider how the ratio of degeneracy, thermal, and rotational energy to the total internal energy of the remnants varies with \qrho.

\paragraph{Energy balance of the entire remnant.} The support against gravity changes from being due mostly to degeneracy pressure at low \qrho\ to having a substantial rotational contribution at $q_\rho\simeq1$.  This is because for highly dissimilar-mass mergers most of the internal energy is locked up within the accretor, which is hardly heated or spun up.  For similar-mass mergers, however, donor material mixes, to some degree, with the entire accretor, causing heating and spin-up throughout the entire remnant.

The total gravitational potential energy of the merger remnant can be described adequately by a constant fraction of $GM_\mrm{tot}^2/\Ra$,
\eqbegin
\frac{-E_\mathrm{pot}}{GM_\mrm{tot}^2/\Ra} = 0.49
\qquad(\pm0.01).
\eqend


From the virial theorem, the internal energy should be related to the potential energy by $3(\langle\gamma\rangle - 1)E_\mathrm{I} = -E_\mathrm{pot}$, where $\langle\gamma\rangle$ is an appropriately averaged equivalent to the adiabatic index.  Since our remnants have cores where the electrons are becoming relativistic, one has $\langle\gamma\rangle$ somewhat smaller than $5/3$, especially for the more massive remnants.  We find that the ratio $E_\mathrm{I}/|E_\mathrm{pot}|$ can be described by,
\eqbegin
\frac{E_\mathrm{I}}{|E_\mathrm{pot}|} = 0.18\frac{M_\mathrm{a}}{M_\odot} + 0.42
\qquad(\pm0.01),
\eqend
which is $\sim$0.5 and $\sim$0.6 for low and high $M_\mathrm{a}$, respectively.  

\begin{figure}
\centering
\includegraphics[width=1.0\columnwidth]{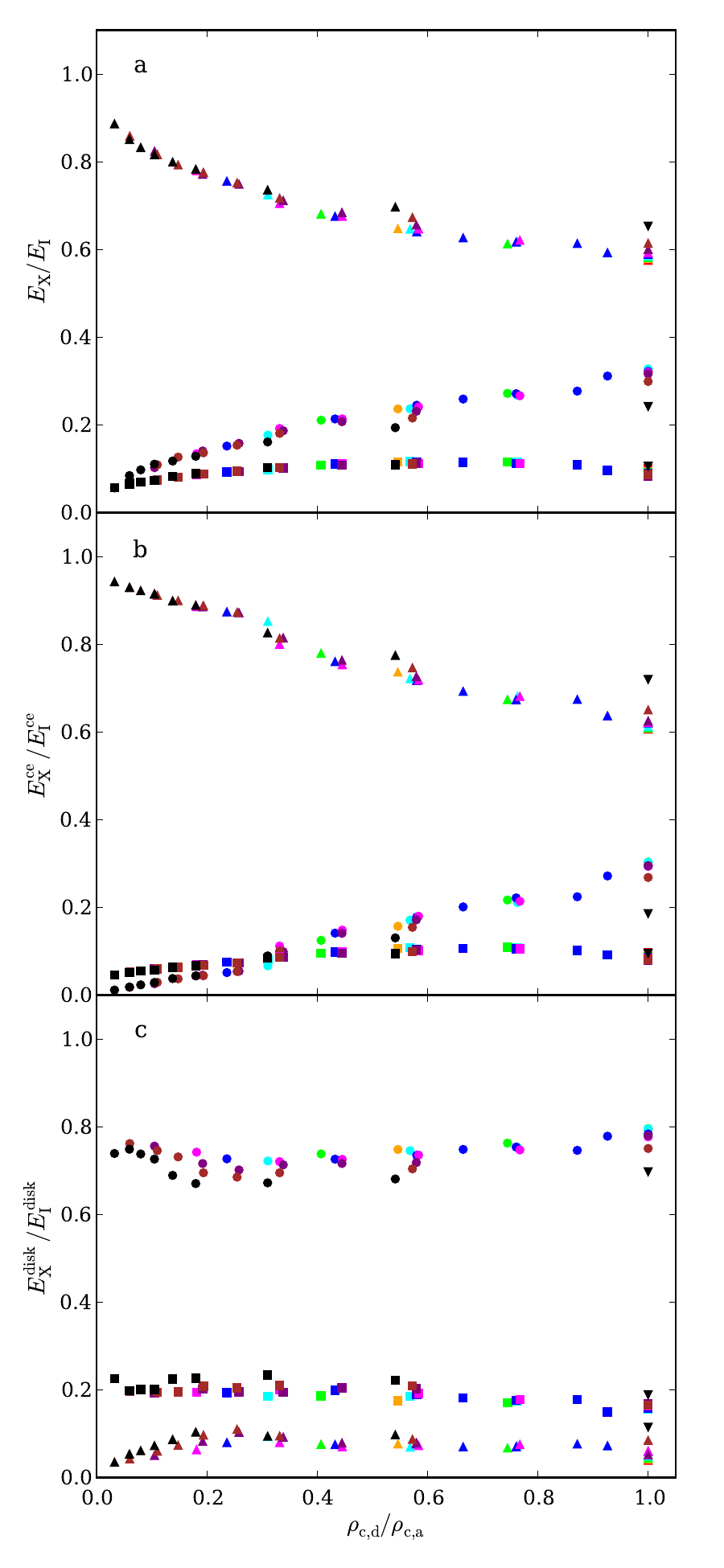}
\caption{Partition of energies in (a) the overall merger remnant, (b) the remnant core plus envelope, and (c) the remnant disk.  In each panel, the fraction of total energy carried in degeneracy (triangles), thermal (squares), and rotational (circles) energy is shown.  Colors represent different accretor masses, encoded as in Fig.~\ref{fig:constacc}.  Triangles represent the 1.0 - 1.0\,\Msun\ merger.}
\label{fig:energytrends}
\end{figure}

The fraction of the internal energy carried by degeneracy and rotation is fairly well described by,
\begin{eqnarray}
\frac{E_\mathrm{rot}}{E_\mathrm{I}} &=& 0.31\qrho^{1/2}
\qquad(\pm0.01),\\
\frac{E_\mathrm{deg}}{E_\mathrm{I}} &=& 0.92 - 0.34\qrho^{1/2}
\qquad(\pm0.02).
\end{eqnarray}
With these, the fraction carried by thermal energy can also be calculated; as shown in Fig.~\ref{fig:energytrends}a, the fraction in thermal energy first increases with increasing \qrho, but turns over at $\qrho\simeq0.7$, decreasing afterwards.  This reflects the competition between increased thermal energy from the two stars mixing, and increased rotational support from the spin-up of the core.  

Overall, for highly dissimilar-mass mergers, the internal energy is partitioned into degeneracy, rotational and thermal energy with a ratio of approximately 8:1:1, reflecting that, as stated above, such mergers are almost entirely supported by degeneracy pressure.  Similar-mass mergers, on the other hand, partition their internal energies with the ratio 6:3:1, i.e., rotational support is significant.  

\paragraph{Energy balance of the core-envelope.} Since the variations with \qrho\ seen for the remnant as a whole are almost entirely due to variations in the core and envelope rather than in the disk, the trends we find for the core-envelope are very similar to those we found above for the entire remnant,
\begin{eqnarray}
\frac{E^\mrm{ce}_\mathrm{rot}}{E^\mrm{ce}_\mathrm{I}} &=& 0.28\qrho
\qquad(\pm0.02), \\
\frac{E^\mrm{ce}_\mathrm{deg}}{E^\mrm{ce}_\mathrm{I}} &=& 0.94 - 0.32\qrho
\qquad(\pm0.02).
\end{eqnarray}
Note the dependency on \qrho, rather than on $\qrho^{1/2}$ as was found for the entire remnant.  See Fig.~\ref{fig:energytrends}b.

\paragraph{Energy balance of the disk.} For the disk, we find very little dependence on \qrho, consistent with the idea that most of the changes in the partitioning of energy have to do with increased mixing between the donor and accretor, which affects the core and envelope much more than the disk.  Averaged over all mergers, we find
\begin{eqnarray}
\frac{E^\mrm{disk}_\mathrm{rot}}{E^\mrm{disk}_\mathrm{I}} &=& 0.74
\qquad(\pm0.03),\\
\frac{E^\mrm{disk}_\mathrm{th}}{E^\mrm{disk}_\mathrm{I}} &=& 0.19
\qquad(\pm0.02),\\
\frac{E^\mrm{disk}_\mathrm{deg}}{E^\mrm{disk}_\mathrm{I}} &=& 0.07
\qquad(\pm0.02).
\end{eqnarray}
Hence, the disk is composed of non-degenerate, primarily rotationally-supported material.  See Fig. \ref{fig:energytrends}c.  (Note that we do not try to define a ratio of internal to potential energy of the disk or core-envelope, since the potential energy of either is not straightforward to determine.)

\subsubsection{Temperature and Thermal Energy}
\label{sssec:thermtrends}

Since heating of the remnant is achieved through shocks and viscous dissipation, the most heavily mixed regions should also be the hottest.  We focus on equatorial thermodynamic values, but consider the rotational axis as well for similar-mass mergers.

\paragraph{The maximum temperature.} We find that the maximum temperature on the equatorial plane, \Tmax, scales with the potential of the accretor (Fig.~\ref{fig:restoftrends}e),
\eqbegin
\frac{k\Tmax}{G\Ma m_\mathrm{p}/\Ra} = 0.20
\qquad(\pm0.03).
\eqend
This scaling is natural in the limit of highly dissimilar-mass merger -- for each nucleon, of order $G\Ma m_\mathrm{p}/\Ra$ is liberated and converted into thermal energy.  The temperature and thermal energy profiles in Fig.~\ref{fig:constacc} show that with increasing {\qrho}, additional thermal energy is deposited into the remnant, but this energy is spread over a larger region, such that the maximum temperature remains roughly the same even as {\qrho} approaches unity.  

For a dissimilar-mass merger, the highest temperature along the rotational axis, \zTmax, is found at the tenuous outer edge of the hot envelope.  It is slightly higher than the maximum temperature found in the equatorial plane.  With increasing \qrho, however, the difference increases noticeably due to the two off-center hot spots found along the rotational axis in similar-mass mergers.  Fitting \zTmax, we find $k\zTmax/(G\Ma m_{\rm p}/\Ra) = 0.24\pm0.03$, though this does not capture the upturn for similar masses well.

All remnants with $q_\rho\gtrsim0.8$ have convectively unstable cores along the equatorial plane.  Artificially mixing these cores to make them isentropic decreases their maximum temperatures by 10 -- 50\% (not shown in Fig.~\ref{fig:restoftrends}, but see the left panel of Fig.~\ref{fig:willitexplode}).  All remnants are stable against convection along the rotational axis.

\paragraph{The density at the point of maximum temperature.}  For dissimilar-mass mergers, the density at the hottest equatorial point, \rhoTmax, depends mostly on the donor (see Fig.~\ref{fig:restoftrends}f).  For $\qrho\lesssim0.5$, we find
\eqbegin
\frac{\rho(T_\mathrm{max})}{\rhocd} = 0.34
\qquad(\pm0.02).
\eqend
This proportionality again suggests that, at least for dissimilar-mass mergers, the donor mixes with the accretor up to a fraction of the central density of the donor, as alluded to earlier.  

At $\qrho\gtrsim0.6$, the dependence becomes less obvious, with \rhoTmax\ varying from $\sim\!25$ -- 90\% of \rhocd.  Since for these density ratios,  the donor material starts to penetrate the central core of the accretor -- and the accretor starts to disrupt as well -- the simple picture of the donor mixing up to a fraction of its own central density may be breaking down.  Furthermore, part of the spread in density reflects that for high \qrho\ the equatorial temperature profiles become nearly flat (see Fig.~\ref{fig:constacc}, left column), thus increasing the sensitivity to noise in the determination of the location (but not the value) of maximum temperature.  This also affects our results for the enclosed mass, \MencTmax, below.

Since for dissimilar-mass mergers, \zTmax\ is located near the tenuous outermost regions of the hot envelope, where particle noise is high, the density at the point of maximum rotational axis temperature, \zrhoTmax\ (plus symbols in Fig. \ref{fig:restoftrends}f), varies wildly between about \rhoTmax\ and one order of magnitude below \rhoTmax.  For similar-mass mergers, \zrhoTmax\ appears to be $\sim\!20$ -- 50\% of {\rhoTmax}.

\paragraph{The mass enclosed within the radius of maximum temperature.}  For dissimilar-mass mergers, the radius of maximum temperature occurs at an enclosed mass of $\MencTmax\simeq\Ma$, while for mergers with $\qrho\gtrsim0.8$, maximum temperature occurs at the center and $\MencTmax\simeq0$ (see Fig.~\ref{fig:restoftrends}g).  For $\qrho\lesssim0.5$, we find
\eqbegin
\frac{\MencTmax}{\Ma} = 1-0.28\qrho
\qquad(\pm 0.01),
\eqend
where the fit's y-intercept is forced to unity.  Note that maximum temperature occurs near the bottom of the mixed zone, which is why \MencTmax\ is substantially smaller than \MMfifty.  The reasons it starts to deviate from a tight trend at $\qrho\simeq0.6$ are the same as those for \rhoTmax: the break-down of the simple mixing picture and the difficulty in determining the location of peak temperature for a broader plateau.

Since $\MencTmax<\Ma$, it may be surprising that \rhoTmax\ is not higher than \rhocd.  This is because the additional thermal and rotational support against gravity reduces the density gradient that would be required if degeneracy pressure were the only source of support.

For dissimilar-mass mergers, \zMencTmax\ is only slightly higher than \MencTmax, since the remnant core is nearly spherically symmetric, apart from the fact that the hot envelope is slightly more extended in the vertical direction.  For similar-mass mergers, the difference increases, reflecting the development of the off-center hot spots, until $\zMencTmax/\Ma\simeq0.25$ for equal-mass mergers.

\paragraph{The mass enclosing half the remnant thermal energy.}  As a more robust measure of where thermal energy is deposited during the merger, we consider the mass enclosing half the remnant thermal energy, \MEthermfifty\ (see Fig.~\ref{fig:restoftrends}h).  We find this is very close to the mass of the accretor,
\eqbegin
\frac{\MEthermfifty}{\Ma}= 1.06
\qquad(\pm 0.04),
\eqend
if the 1.0 - 1.0 {\Msun} merger is neglected.  One sees turnovers at the extremes, for $\qrho\lesssim0.2$ and $\qrho\gtrsim0.8$.  The former likely is because thermal energy is deposited into a narrow strip right on the surface of the accretor, while the latter is probably due to the disruption of the accretor.

While smaller than \MMfifty, \MEthermfifty\ is always larger than \MencTmax.  This reflects that high density degenerate material has lower specific heat, so that for the same energy per unit mass the temperature is higher (see Fig.~\ref{fig:constacc}).  

\paragraph{The width of the remnant thermal energy.}  The mass enclosed between the 25$^{\rm th}$ and 75$^{\rm th}$ percentiles of thermal energy, $\MEthermthick = M_{\rm enc}(\frac{3}{4}E_\mathrm{th}) - M_{\rm enc}(\frac{1}{4}E_\mathrm{th})$, is a measure of the extent of the remnant that has been heated (see Fig.~\ref{fig:restoftrends}i).  Ignoring the 1.0 - 1.0 \Msun\ merger, it can be fit by,
\eqbegin
\frac{\MEthermthick}{\Ma} = 0.11+0.94\qrho
\qquad(\pm0.03).
\eqend
Here, we did not force the y-intercept to go to zero, which is expected physically but gives a substantially poorer trend.

\subsubsection{Angular Velocity and Rotational Energy}
\label{sssec:rottrends}

For a dissimilar-mass merger, the donor carries most of the angular momentum.  As a result, the hot envelope contains more angular momentum and features higher angular velocities than the core, since the envelope is where most of the accreted donor material resides.  Spin-up of the accretor is accomplished through shocks, $PdV$ work and shearing forces.  For a similar-mass merger, the two stars carry similar amounts of angular momentum and thoroughly mix.  Conservation of angular momentum then implies that the entire remnant rotates rapidly.

\paragraph{The maximum angular velocity.}  On the equatorial plane, the highest angular velocity, \Omegamax, scales linearly with the orbital angular velocity of the pre-merger binary, $\Omegaorb = 2\pi/P_\mrm{orb}$ (see Fig.~\ref{fig:restoftrends}j),
\eqbegin
\frac{\Omegamax}{\Omegaorb} = 3.8
\qquad(\pm0.6).
\eqend

\paragraph{The ratio of core-envelope to total angular momentum.} For more similar-mass mergers, more angular momentum is deposited in the accretor and ends up in the core and envelope (see Fig.~\ref{fig:restoftrends}k).  The ratio of core-envelope to total angular momentum, \Lrat, is approximately,
\eqbegin
\frac{L_\mrm{ce}}{L_\mrm{tot}} = 0.70\qrho
\qquad(\pm 0.03),
\eqend
where we fit only for $\qrho > 0.25$ and ignore the 1.0 - 1.0 {\Msun} merger.  For $\qrho\lesssim0.25$, the trend becomes shallower, resulting in a non-zero intercept.  This suggests that even in cases where the donor has negligible mass some angular momentum is transferred to the accretor.

\paragraph{The mass enclosed inside the radius of maximum angular velocity.}  
For dissimilar-mass mergers, both \MencOmax\ and \MencTmax\ are about equal to \Ma, with \MencOmax\ slightly larger than \MencTmax: for $\qrho\lesssim0.55$, $\MencOmax/\Ma\,=\,1.05\pm0.06$.  This is consistent with the idea that the hottest and most spun-up regions are those where the donor mixed most strongly with the accretor.  For $\qrho\simeq0.6$, the off-center angular velocity peak is replaced by a plateau, and \MencOmax\ becomes ill-defined; for even larger \qrho, the highest velocities occur in the center, and $\MencOmax\simeq0$.  See Fig.~\ref{fig:restoftrends}l.


\paragraph{The mass enclosing half the remnant rotational energy.} Like for the thermal energy, for very dissimilar masses, the mass enclosing half the rotational energy, \MErotfifty, is similar to the accretor mass (see Fig.~\ref{fig:restoftrends}m).  For $\qrho\lesssim0.8$, we find
\eqbegin
\frac{\MErotfifty}{\Ma} = 1.12+0.27\qrho^{1/2}
\qquad(\pm0.01)
\eqend
Note that unlike {\MencOmax}, \MErotfifty\ continues to increase with \qrho\ (except for exactly equal-mass mergers), a consequence of particles with lower angular velocity but large lever arm that carry substantial rotational energy (see Fig.~\ref{fig:constacc}).  Near $\qrho\simeq0.8$, the trend breaks as both stars are significantly disrupted.  However, even exactly equal-mass mergers have more of their rotational energy stored in the outskirts (otherwise one would have $\MErotfifty/\Ma\simeq1$).  

\paragraph{The width of the remnant rotational energy.}  We measure the extent to which the remnant is affected by spin-up through the difference between the masses enclosing 25 and 75\% of the rotational energy, $\MErotthick = M_{\rm enc}(\frac{3}{4}E_\mathrm{rot}) - M_{\rm enc}(\frac{1}{4}E_\mathrm{rot})$ (see Fig.~\ref{fig:restoftrends}n).  Ignoring the 1.0 - 1.0 {\Msun} merger, it is well-described by,
\eqbegin
\frac{\MErotthick}{\Ma} = 0.12+0.77\qrho
\qquad(\pm0.03).
\eqend
Like for the thermal energy, one sees that for more similar-mass mergers, rotational energy is spread more widely throughout the remnant.

\subsection{A Qualitative Picture of the Merger}
\label{ssec:qualitative}

From our empirical results above, a qualitative picture of a merger emerges.  A dissimilar-mass merger has the donor overflowing its Roche lobe and forming an accretion stream.  This stream mixes with the accretor up to approximately the central density of the donor pre-merger, \rhocd.  Those layers of the accretor that are denser than \rhocd\ are hardly affected, and form the cold core of the merger remnant, while the mixed material will form a partly-thermally supported outer envelope, which somewhat compresses the core, as well as a rotationally supported disk.

At $\qrho\gtrsim0.6$, the above picture begins to break down, as portions of the donor start to penetrate to the center of the accretor.  This results in substantial heating and spin-up of the central core: the merger becomes a similar-mass merger.  As the masses become more similar, the distinction between donor and accretor is lost and both stars disrupt and form accretion streams.  For all $\qrho\gtrsim0.6$, the remnants are similar: a large, ellipsoidal and partly rotationally supported hot core with two hotspots off the equatorial plane, surrounded by a small, hot disk.

For all mergers, the maximum temperature reached by dissipation of orbital energy is proportional to the accretor's gravitational potential energy.  For increasing \qrho, the maximum temperature remains similar, but the region over which the thermal energy is deposited widens.  The density at maximum temperature is of the same order of magnitude as the central density of the donor, consistent with the mixing picture discussed above.  The latter no longer holds for $\qrho\gtrsim0.6$, when the entire remnant is mixed and heated.

For a dissimilar-mass merger, the angular momentum remains in the outer regions, since most of it was originally carried by the donor.  Angular momentum can be transferred between regions through shocks, $PdV$ work or shearing forces, all of which becomes increasingly important as the donor penetrates deeper.  As a result, with increasing \qrho, the remnant core is spun up further.  Where both WDs disrupt, leading to colliding accretion streams, even the densest regions of the remnant have high rotational velocities.

\section{Variation of Merger Parameters and Robustness of Results}
\label{sec:variation}

In our parameter space study, we focused on the effects of varying the masses of the two WDs, fixing the initial separation {\azero}, merger completion time criterion, and WD composition.  To determine how robust our results are, we ran simulations varying these assumptions.

\subsection{Changing the Composition}

\begin{figure}
\centering
\includegraphics[angle=0,width=1.0\columnwidth]{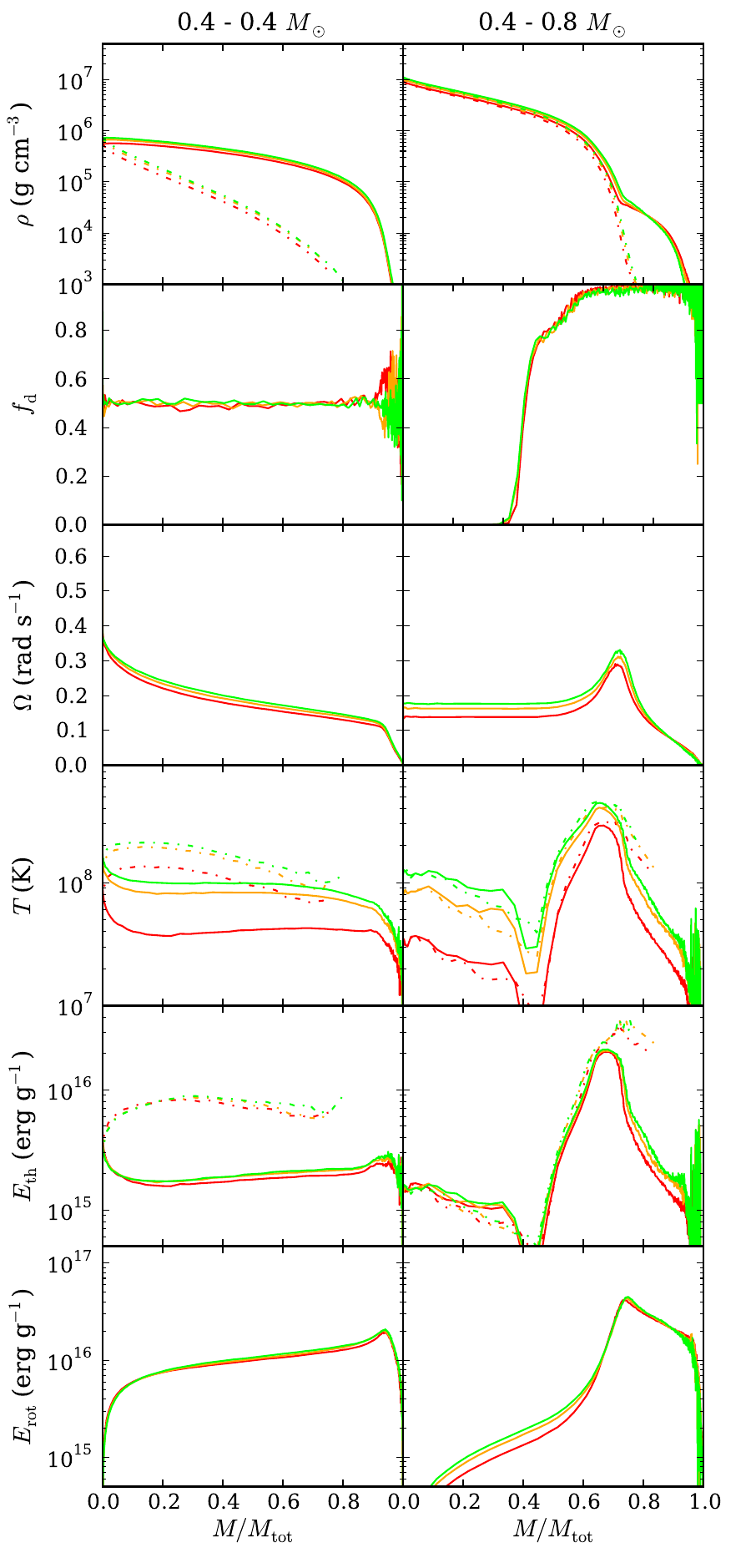}
\caption{As Fig.~\ref{fig:constacc}, but for 0.4 - 0.4 {\Msun} (left) and 0.4 - 0.8 {\Msun} (right) mergers with different compositions: pure $^4$He (red), CO (orange), and pure $^{24}$Mg (lime).  Dash-dotted lines represent profiles along the rotational axis rather than the equatorial plane.}
\label{fig:compcomp}
\end{figure}

Ignoring fusion, WD mergers should be insensitive to changes in composition, since the dominant electron degeneracy pressure only depends on the mean molecular weight per electron, which is close to $\mu_e\simeq2$ for all likely compositions.  To confirm this, we ran simulations assuming pure helium and pure magnesium for an equal-mass case (0.4 - 0.4\,\Msun) and an dissimilar one (0.4 - 0.8\,\Msun).  The results are shown in Figure \ref{fig:compcomp}.  One sees that most quantities indeed have very similar profiles.

The set of profiles showing most variation are those of the temperature.  These have similar shape, but different normalization.  Since the thermal energy curves are very similar, it is clear that this reflects differences in heat capacity, which does depend on composition: for He composition, there are more (non-degenerate) ions than for our standard CO mixture, boosting the heat capacity (and thus lowering the temperature for given thermal energy), while for Mg composition, there are fewer, lowering the heat capacity (and increasing the temperature).  As a result, the maximum equatorial plane temperatures for the 0.4 - 0.4\,\Msun\ simulations are 0.95, 1.48 and $1.68\times10^8$ K for He, CO, and Mg, respectively, while for the 0.4 - 0.8\,\Msun\ simulations, they are 2.92, 4.05, and $4.47\times10^8\,$K.


The smaller differences seen for the other profiles reflect small differences in initial conditions.  All WDs are constructed assuming $T=5\times10^6\,$K throughout, which implies more thermal energy for higher heat capacity.  As a result, the relaxed He and Mg WDs are slightly larger and smaller, respectively, than the CO WD.  These slight differences in radius translate into differences in initial separation, which in turn cause small differences in the angular velocity and rotational energy curves.

\subsection{Varying the Initial Binary Separation}
\label{ssec:varyingazero}

\begin{figure}
\centering
\includegraphics[angle=0,width=1.0\columnwidth]{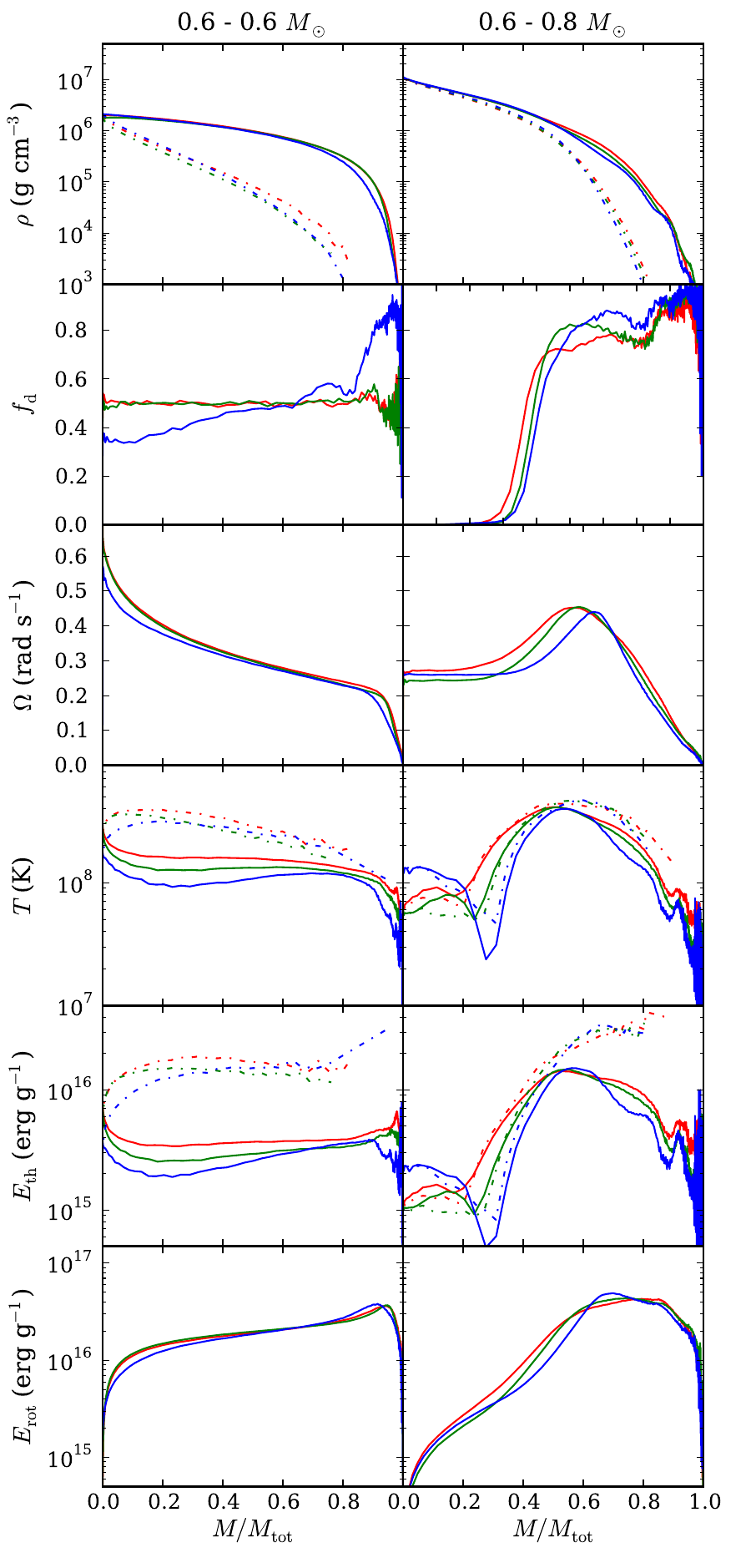}
\caption{As Fig.~\ref{fig:compcomp}, but for 0.6 - 0.6 {\Msun} (left) and 0.6 - 0.8 {\Msun} (right) mergers with varying initial orbital separation: 0.9 (red), 1.0 (green), and 1.1 (blue) times the value used for the parameter space study.}
\label{fig:distcomp}
\end{figure}

For our simulations, we chose an initial orbital separation {\azero} for which a co-rotating donor would fill its Roche lobe.  Since our (non-rotating) WDs are equilibrated in isolation, once the simulation starts they immediately begin to adjust to the tides and hence disrupt quickly.  Ideally, one would allow them to adjust to the binary potential and start mass transfer properly.  For non-synchronous rotation, however, this is not straightforward (see Sec.~\ref{ssec:compwithothers}).  Nevertheless, we try to get a sense of the influence of this by running simulations for two cases -- 0.6 - 0.6\,\Msun\ and 0.6 - 0.8\,\Msun\ -- with \azero\ increased and decreased by 10\% (see Fig.~\ref{fig:distcomp}).

Our default simulations were considered complete at 6 orbits of the initial binary.  For runs where \azero\ was changed, we used 2.5\% non-axisymmetry, and a requirement for the density to be highest at the remnant's center, as the completion criteria\footnote{The 2.5\% non-axisymmetry convergence time is 312 s (6.6 orbits) for our default 0.6 - 0.6 {\Msun} run, and 250 s (5.2 orbits) for our default 0.6 - 0.8 {\Msun} run.}.  Not surprisingly, runs with larger {\azero} needed longer to achieve these criteria: with a 10\% increase, the 0.6 - 0.8 {\Msun} merger required 861 seconds, or $\sim\!15$ orbits, to complete, while the one with a 10\% decrease required 230 s, or $\sim\!5.5$ orbits.  Similarly, the 0.6 - 0.6 {\Msun} merger with a 10\% increase in {\azero} required $\sim\!13$ orbits (725 s) to complete, compared to the 7 orbits of a 10\% decrease (283 s).  In both cases, the increase simply reflects that it takes longer for the donor to be disrupted fully if \azero\ is increased\footnote{Of course, if placed far enough, the binary does not merge.  For a 0.6 - 0.8 {\Msun} binary, no mass transfer occurred within 500 s if \azero\ was increased by 20\%.}.  For instance, for the 0.6 - 0.8 {\Msun} binary with increased separation, it took almost a dozen orbits before full disruption, while at the standard separation disruption occurred after just 1.5 orbits.  This feature is of particular interest because for mergers of synchronously rotating WDs \cite{dan+11,dan+12} and \cite{rask+12} all note almost immediate disruption of the donor when approximate initial conditions are used, and much delayed disruption for more accurate initial conditions (up to $\sim\!30$ orbits; see Sec.~\ref{ssec:compwithothers}).

We find that the density profiles of the merger remnants are remarkably insensitive to varying \azero\ (\rhoc\ changing by $\lesssim2$\% for the 0.6 - 0.8 {\Msun} merger, and $\sim20$\% for the 0.6 - 0.6 {\Msun} merger), and show substantial systematic changes only in the outer regions.  The latter can be understood from the mixing and rotational profiles, where one sees that with increasing \azero, donor material is mixed less deeply into the accretor, and rotational energy is shifted outward, causing the rotational frequency to peak at lower values and larger radii.  This reflects the increase in angular momentum with increasing \azero, which creates a more rotationally supported remnant (for both our systems, a 10\% increase (decrease) in \azero\ results in a 5\% increase (decrease) in angular momentum).  In the 0.6 - 0.8\,\Msun\ merger, the decreased mixing causes the accretor to be spun up less, thus lowering the rotational energy of the core, and narrowing the thermal energy plateau.  These effects are also seen in the similar-mass case, where the center of the remnant receives less rotational support and becomes denser with increasing separation, and the mixing becomes less uniform.

Qualitatively, with increasing \azero, the properties change in a way that is similar to the changes seen with decreasing \qrho, i.e., similar to mergers with more dissimilar mass: reduced mixing, larger disks and less core rotational support, and shifts in the thermal and rotational energies toward larger radii.  The converse is also true, decreasing \azero\ has similar effects as increasing \qrho, i.e., the mergers become similar to those with more equal masses.  The changes are substantial at times: e.g. with a 10\% increase in {\azero} for the 0.6 - 0.6 {\Msun} remnant, the maximum equatorial temperature is reduced by 40\%, while the corresponding density increases by 25\% (for the rotational axis hotspots, the values are a 13\% and 30\% reduction, respectively), and the mass of the disk increases by 65\%.  Similar, though far less extreme, changes are seen for the properties of the 0.6 - 0.8 {\Msun} remnant.  All this makes {\azero} one of the parameters our mergers are most sensitive to.

\subsection{Synchronization}
\label{ssec:synchronization}

\begin{figure}
\centering
\includegraphics[angle=0,width=1.0\columnwidth]{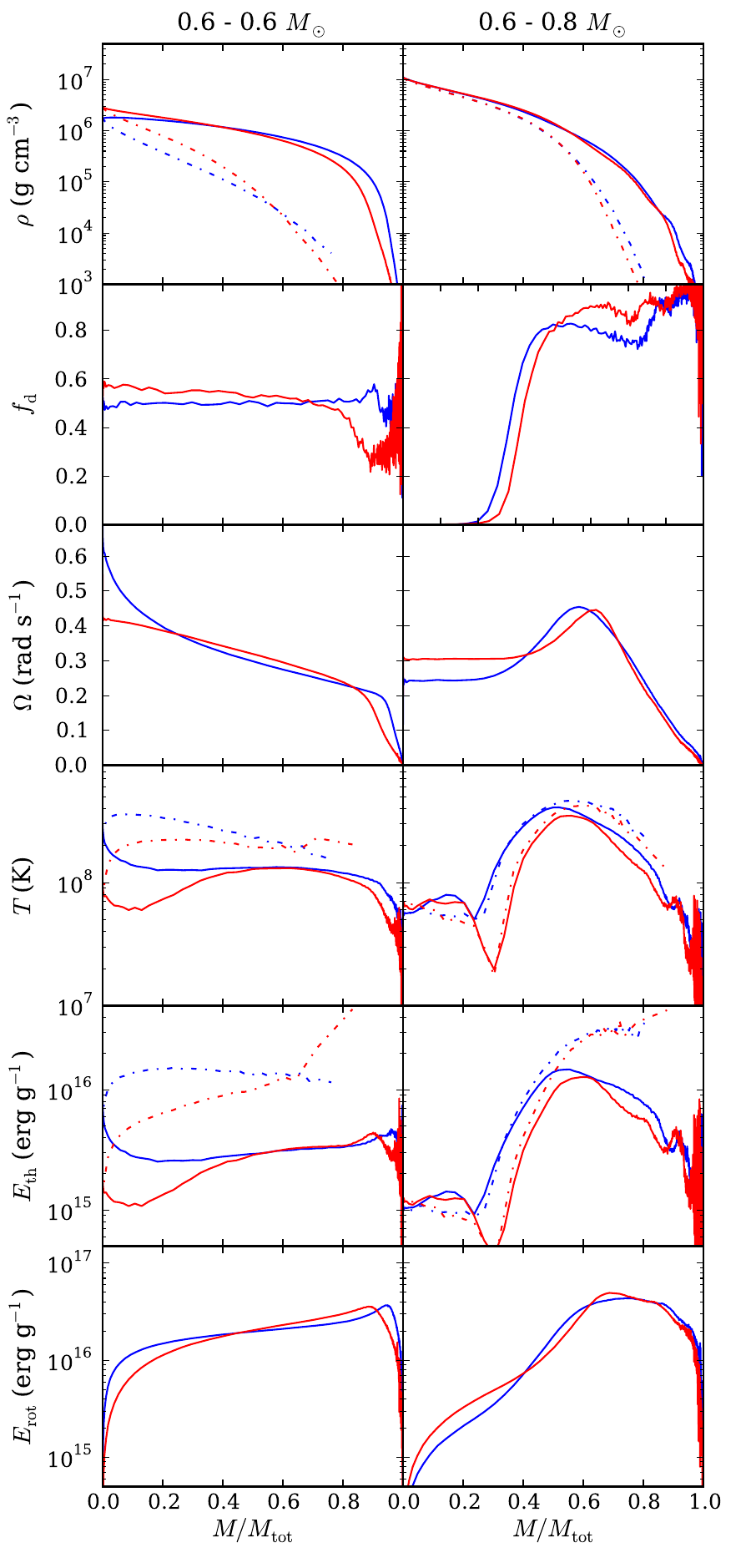}
\caption{As Fig.~\ref{fig:compcomp}, but for 0.6 - 0.6 {\Msun} (left) and 0.6 - 0.8 {\Msun} (right) mergers, comparing our default, irrotational case (blue) with that assuming synchronous rotation (red).}
\label{fig:synchcomp}
\end{figure}

In our simulations, the WDs have zero spin, i.e., we assume that tidal dissipation is too weak to synchronize their rotation.  Whether or not this is correct is currently unknown, but to see what the effect could be, we ran simulations assuming synchronized rotation for 0.6 - 0.6\,\Msun\ and 0.6 - 0.8\,\Msun\ binaries (see Fig.~\ref{fig:synchcomp}).  We used approximate initial conditions for these systems, identical to those of Sec. \ref{ssec:initcond} except that the stars rotate at the orbital angular frequency.

As in Sec.~\ref{ssec:varyingazero}, our unsynchronized runs are from our parameter space study, and use 6 orbits as their completion criterion, while our synchronized runs use 2.5\% non-axisymmetry, and a requirement for the density to be highest at the remnant's center.  Completion occurred at 407 s ($\sim$8.5 orbits) for the synchronized 0.6 - 0.6 {\Msun} simulation, and 314 s ($\sim$6.5) orbits for the synchronized 0.6 - 0.8 {\Msun}.

The asynchronous and synchronous mergers differ mostly in the amount of heating and spin-up.  This has two causes.  First, for the synchronized binary, the total amount of angular momentum is about 10\% larger, and high angular momentum material has more difficulty penetrating the accretor, as is evident in the 0.6 - 0.8 {\Msun} mixing profile (because of the larger amount of angular momentum, the mergers also take about 1.5 orbits longer to achieve 2.5\% non-axisymmetry).  Second, in a synchronized binary, the donor and accretor have much less differential rotation with respect to each other, leading to much less spin-up and heating.

Both effects are largest for the equal-mass case.  In particular, in a synchronized, equal-mass binary contact can occur without any friction, while in an unsynchronized one it involves shocks at the full orbital velocity.  In consequence, for the synchronized case, rotational support is weaker in the center and stronger in the outskirts, causing the central density of our 0.6 - 0.6 {\Msun} remnant to increase by $\sim70$\% and the disk mass to increase by a factor of 2.  Furthermore, while for the non-synchronized case, the maximum temperature along the equatorial plane was found in the center, for the synchronized case it is found in the outskirts, and is more than a factor two lower ($1.3\times10^8\,$K instead of $2.9\times10^8\,$K).  The hotspots on the rotational axis also have much reduced temperature, $2.3\times 10^8$\,K instead of $3.6 \times 10^8$\,K.  

For the dissimilar-mass merger, the effects of synchronization are less dramatic: the accretor still spins up substantially, and rotational and thermal energy are deposited in roughly the same way.  The main difference is that the synchronized case has slightly less mixing, causing a drop in total thermal energy and maximum temperature (from $4.1\times10^8$ K to $3.5\times10^8$ K on the equatorial plane, and from $4.6\times10^8$ K to $4.3\times10^8$ K on the rotational axis).


\subsection{Running the Simulation Longer}
\label{ssec:runninglonger}

\begin{figure}
\centering
\includegraphics[angle=0,width=1.0\columnwidth]{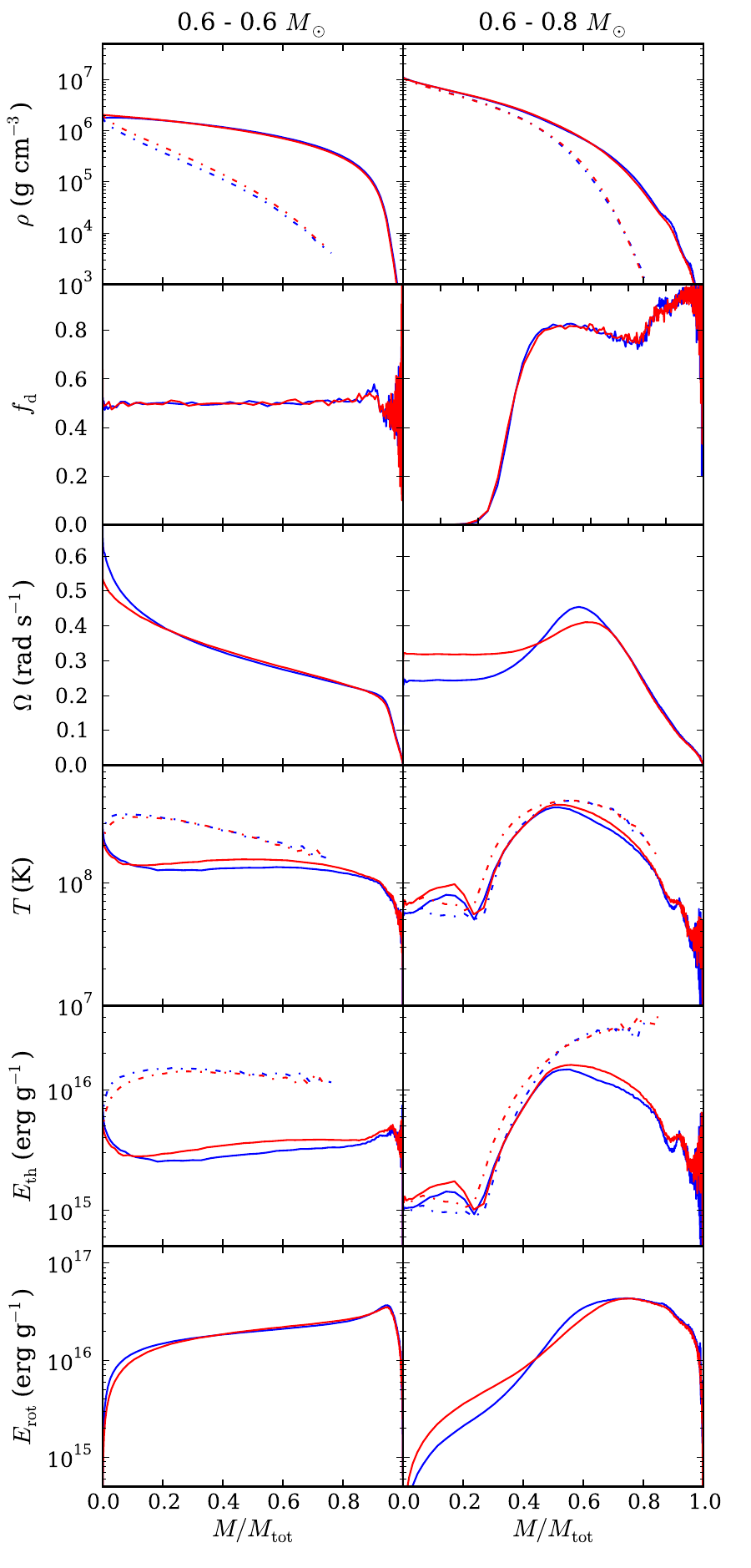}
\caption{As Fig.~\ref{fig:compcomp}, but for 0.6 - 0.6 {\Msun} (left) and 0.6 - 0.8 {\Msun} (right) mergers, comparing properties for our default simulation time of 6 initial orbital periods (blue) with those obtained after 8 orbital periods (red).}
\label{fig:timecomp}
\end{figure}

We considered our mergers completed after 6 orbits, since in that time they on average had reached our convergence criterion of 2.5\% non-axisymmetry (see Sec.~\ref{ssec:mergercomplete}).  To test the robustness of our results, we also determined properties attained after 8 orbits.  In Fig.~\ref{fig:timecomp}, we compare the 6 and 8 orbit results for 0.6 - 0.6\,\Msun\ and 0.6 - 0.8\,\Msun\ binaries.

We find our mergers show little evolution between 6 and 8 orbits, with the largest changes seen for the angular velocity profiles.  For the 0.6 - 0.8 {\Msun} merger, the rigidly rotating core sped up and the off-center peak decreased in height and moved out, while for the 0.6 - 0.6 {\Msun} merger, the center spun down and the rotational profile became flatter.  In the dissimilar-mass merger, \Tmax\ and \rhoTmax\ changed by $\lesssim\!5$\% and the density and temperature structures nearly overlap, while in the equal-mass merger \Tmax\ and \rhoTmax\ changed by $\sim\!20$\% ($2.9$ to $2.3\times10^8\,$K and $1.7$ to $2.0\times10^6\,\gcc$), reflecting an increase in central density and a shifting of the temperature profile, with temperature decreasing in the center but increasing elsewhere.  The evolution of all properties is consistent with viscous evolution -- expected to follow the merger proper -- with the core driven into rigid rotation, and angular momentum transferred outward to the disk.  In the dissimilar-mass merger, the net effect is spin-up of the core and spin-down of the envelope, while in the equal-mass case it is the reverse.  Of course, in the process, rotational energy is turned into thermal energy, heating the remnants.

One curious aspect for equal-mass mergers is the evolution of the off-center hot spots (Fig.~\ref{fig:mergersampling2}).  Over time, these broaden parallel to the equatorial plane, yet become narrower along the rotational axis.  As a result, the hourglass shape is lost, and the center of the remnant stops being one of the hottest point in the system.  After 8 orbits, the system resembles more closely what we find for typical similar-mass mergers, which have more pancake-shaped hot spots flanking a colder, denser region on the equatorial plane.

Comparing more broadly the 6 and 8-orbit results, we find changes at the $\sim\!5$\% level.  The trends presented in Sec.~\ref{ssec:mergertrends} continue to hold to within $\sim\!10$\% except for \hxy\ (which becomes $\hxy/h_{\rm a} = 0.98 + 0.74q_\rho^2~(\pm0.07)$), the fraction of disk energy in degeneracy energy ($E_\mathrm{deg,disk}/E_\mathrm{I,disk} = 0.07~(\pm0.02)$), mass enclosed within the radius of maximum temperature ($\MencTmax/\Ma = 1-0.21q_\rho~(\pm 0.01)$), maximum rotational frequency ($\Omegamax/\Omegaorb = 3.4~(\pm0.5)$) and the widths of the regions in which thermal and rotational energy are deposited ($\MEthermthick/\Ma = 0.13 + 0.87q_\rho ~(\pm0.03)$; $\MErotthick/\Ma = 0.15 + 0.70q_\rho ~(\pm0.02)$).  Changes to these trends are consistent with the viscous evolution described above: the remnant is beginning to spin down, lose its rotational support, and energy is being redistributed.

\subsection{Viscosity Prescription}
\label{ssec:viscprescrip}

\begin{figure}
\centering
\includegraphics[angle=0,width=1.0\columnwidth]{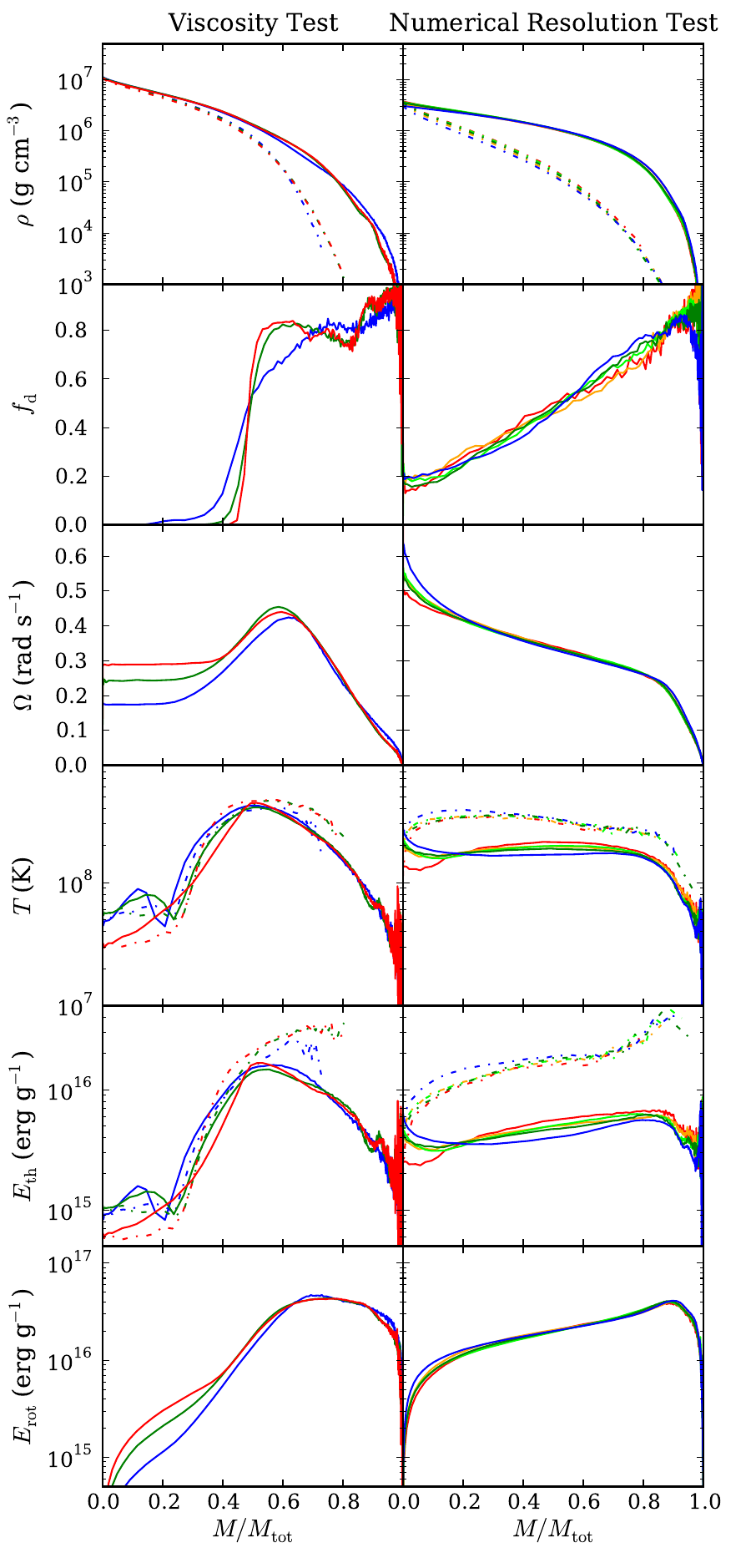}
\caption{As Fig.~\ref{fig:compcomp}, but comparing simulations of a 0.6 - 0.8\,\Msun\ merger with different viscosity prescriptions (left) and simulations of a 0.625 - 0.65\,\Msun\ merger with different numbers of particles (right).  For the viscosity, we compare fixed low viscosity ($\alpha=0.05$, $\beta=0.1$; blue), standard variable viscosity (green), and fixed high viscosity ($\alpha=1.0$, $\beta=2.0$; red).  For particle numbers, we show simulations at one quarter (red), half (orange), and double (blue) the default number of particles, as well as the default simulation (lime) and a rerun of the default simulation (green) to determine the effect of order-of-execution differences, round-off errors and other such numerical effects.}
\label{fig:viscnumcomp}
\end{figure}

The addition of artificial viscosity is required in SPH to accurately capture shocks, but no consensus exists on how best to implement it.  We ran two additional simulations of a 0.6 - 0.8\,\Msun\ merger to check the robustness of our results with respect to changes in the viscosity, one with small and one with large artificial viscosity (fixed $(\alpha,\beta)=(0.05,0.1)$ and $(1,2)$, respectively).  As before, these additional runs use 2.5\% non-axisymmetry and a requirement for the density to be highest at the remnant's center as their completion criteria, and the low viscosity run completed at 198 s while the high completed at 242 s.  Here, we expect that low values of $\alpha$ will lead to large particle noise and inaccurate shock capturing, while high values result in large viscous heating and rapid loss of differential rotation.  Our results confirm this (Fig.~\ref{fig:viscnumcomp}): the simulation with low artificial viscosity leads to a remnant with stronger differential rotation, with the disk carrying 34\% more rotational energy (and the remnant 33\% less) than in the standard variable $\alpha$ simulation.  Lower viscosity also leads to greater mixing of donor and accretor material, reflecting the stronger diffusion associated with the larger particle noise inherent to low viscosity.

Aside from the mixing and spin-up, the results for the three different viscosity prescriptions do not differ greatly.  While one might have expected greater dissipation of rotational into thermal energy for higher viscosity, the maximum temperatures and rotation rates vary by $\lesssim\!10$\%, and the thermal and rotational profiles are quite similar.  The density profiles are virtually identical except near the outer parts, where the low viscosity simulation leaves matter with greater rotational support.

\subsection{Spurious Heating}
\label{ssec:spheat}

As discussed in Sec.~\ref{ssec:sphcode} and throughout Sec.~\ref{sec:results}, noise combined with a pressure floor in the equation of state lead to small increases in internal energy.  While this energy has a negligible effect on most remnant properties, in the most degenerate regions of the remnant it can cause significant temperature increases.  Here, we discuss the extent to which spurious heating affects our results.

\begin{figure}
\centering
\includegraphics[angle=0,width=1.0\columnwidth]{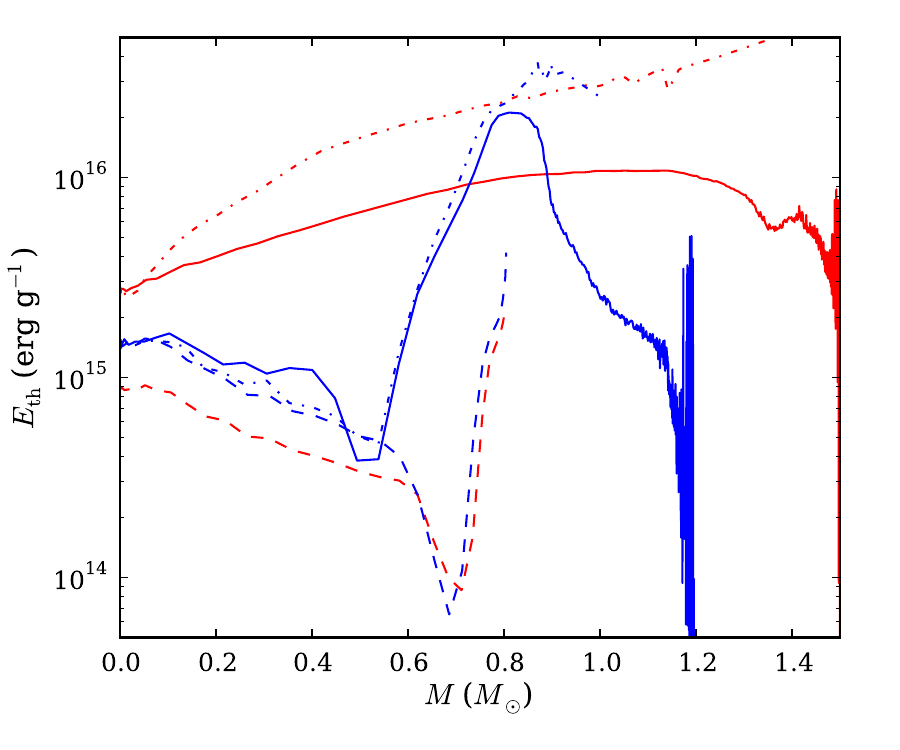}
\caption{Specific thermal energy as a function of enclosed mass for a 0.4 - 0.8 {\Msun} (blue) and a 0.7 - 0.8 {\Msun} merger (red), shown both along the equatorial plane (solid curves) and along the rotational axis (dot-dashed).  Also shown are the (spherical) profiles found for an isolated 0.8\,\Msun\ white dwarf (dashed) simulated using the same parameters, and for the same completion times (six initial orbital periods, equivalent to 489 s and 224 s).  Spurious heating is estimated to be responsible for nearly all the thermal energy in the core of the 0.4 - 0.8\,\Msun\ remnant, and for about one third in the core of the 0.7 - 0.8\,\Msun\ remnant.  It is not important in regions heated by interaction.  (The ``hook'' in the outer layers of the white dwarf profile reflects the high initial temperature chosen; in a merger, this is erased by the interaction.)}
\label{fig:heating}
\end{figure}

As a comparison for the spurious heating seen in some of the simulations, we relaxed a 0.8\,\Msun\ isolated white dwarf for $489\,$s longer than the standard 81\,s we used for relaxing single stars.  While the total energy of the WD (potential, degeneracy and thermal energy combined) increased by $\sim\!1$\% of the original total energy over the additional period of time, the change in thermal energy was enough to raise the central WD temperature from $1.2\times 10^7$ K (increased from $5\times 10^6$ K due to particle noise) to $1.2\times 10^8\,$K.  In Fig.~\ref{fig:heating}, we compare the thermal energy profile of this isolated 0.8 {\Msun} WD with those of a 0.4 - 0.8\,\Msun\ and a 0.7 - 0.8\,\Msun\ merger, with the times for the isolated WD taken at 489 and 224\,s (the mergers' respective completion times) longer than the standard 81\,s.  The total thermal energy generated in the WD at an additional 489\,s is $\sim$10\% of the thermal energy generated in a 0.4 - 0.8\,\Msun\ merger.\footnote{A $\sim$10\% increase in thermal energy corresponds to $\sim\!1$\% increase in the overall energy of the 0.4 - 0.8\,\Msun\ remnant, somewhat larger than the typical $\sim\!0.3$\% level at which Gasoline conserves total energy in our simulations.  We find that similar-mass simulations tend to lose total energy at the 0.05\% level, while some of the low {\qrho} mergers gain more than 1\% in total energy due to spurious heating.}  Indeed, given how well the specific thermal energy profiles match in the interior, it is clear that spurious heating dominates there.  

Spurious heating is much less important for the 0.7 - 0.8\,\Msun\ merger, since its core has mixed to much greater extent and less time was needed for the merger to complete.  The total thermal energy generated in the isolated WD at 224\,s is only $\sim\!3$\% of the thermal energy generated during merger, and even at the very center spurious heating contributes $\sim\!35$\% (rather than nearly all) of the thermal energy.  As a result, the central temperature of the merger, $1.4 \times 10^8\,$K, is substantially higher than that of the isolated WD, $7.8 \times 10^7\,$K.  It is important to note this comparison overestimates spurious heating, since the isolated WD will have had many more particles that dip below the Fermi energy than the much hotter merger remnant core. 


Overall, we conclude that spurious heating is present, but is recognized fairly easily and does not influence our conclusions.  In particular, its effects on remnants should be small in both high-density regions with $T\gtrsim3\times10^8$\,K and in lower-density $\lesssim\!10^6\,\gcc$ regions.  Other simulations may suffer from spurious heating as well.  In this respect, it is intriguing that our equatorial temperature curves for a 0.6 - 0.8\,\Msun\ merger are a good match those of \citetalias{loreig09}, even in the center (Fig.~\ref{fig:compwithloreig}; Sec.~\ref{ssec:compwithloreig}).


\subsection{Resolution}
\label{ssec:restest}


To determine whether or not the numerical resolution matters for our results, we ran three additional simulations of a 0.625 - 0.65 {\Msun} merger, with roughly a quarter, half, and double the number of particles (63,736, 127,525 and 510,047, respectively), corresponding to 0.63, 0.79 and 1.26 times the SPH smoothing length (resolution) we normally use.  We also ran a second simulation with the same number of particles (255,035 particles).  Here, we chose 0.625 - 0.65\,\Msun\ to see if numerical resolution has any effect on whether a merger is ``similar-mass''.  All simulations were considered complete at 6 orbits of the initial binary, though we also checked the 2.5\% non-axisymmetry convergence times.

From Fig.~\ref{fig:viscnumcomp}, one sees that the two runs using the same number of particles -- and identical initial conditions and the same version of Gasoline -- still give slightly different results.  This is due to the inherent non-linear nature of fluid dynamics, coupled with small, random perturbations, e.g., from differences in the order of force addition in parallel processing, round-off errors and slight inconsistencies in converting thermal energy to temperature.  Overall, merger remnant properties change by $\sim3.5$\% between the two runs.  The most prominent differences are seen in properties determined from low numbers of particles, such as {\Tc} (varies by $\sim10$\%), and those involving finding maxima of temperature plateaus, such as \zrhoTmax\ (40\%) and $\rho(T^\mathrm{cv}_\mathrm{max})$ (a factor of 3 -- in one case convection shifts the temperature maximum off-center).


The differences for different resolutions are larger.  While to first order, the equatorial density and mixing profiles are very similar, there is a systematic $\sim\!20$\% drop in the equatorial density -- $\sim\!25$\% in the rotational axis density -- near the center of the remnant with increasing numerical resolution (top right panel of Fig.~\ref{fig:viscnumcomp}).  The angular velocity and rotational energy profiles are again very similar, except in the central regions, where there is a $\sim\!20$\% increase in \Omegamax.  For the temperature the effects are larger: with higher resolutions, most of the equatorial plane is colder, with a $\sim20\!$\% drop in the value of the temperature plateau near $M/M_\mrm{tot}\,=\,0.5$.  The temperatures along the rotational axis, however, increase with increasing resolution, by $\sim\!10$\% across the range of resolutions, as does the upturn in equatorial temperature near the center of the remnant, by $\sim\!50$\% ($\sim\!30$\% if we do not include the lowest resolution run).  The latter effects are due to increasing prominence of the off-center hotspots at higher resolutions, which also tend to look more hourglass-shaped.  Indeed, for our lowest resolution, the densest material in the two stars remains relatively cold throughout the entire merger, resembling the synchronized systems described in Sec.~\ref{ssec:synchronization}.  Finally, we find that if we do not consider the lowest resolution run, the disk half-mass radius varies by 4\%, angular velocity at the half-mass radius varies by 4\%, and the core-envelope mass changes by 3\%.  This is similar to the results of the resolution tests of \cite{rask+12}.


The 2.5\% non-axisymmetry convergence times for the half and double-particle number runs are within 14 s of the 275 s non-axisymmetry convergence time of the default run, a small difference that implies a negligible amount of post-merger evolution.  Only the quarter-particle number run deviated substantially, converging 57 s earlier.  This may simply reflect the smaller number of particles in the disk, where the system is most asymmetric.

We stress that even though the order-of-magnitude change in particle number (factor of two change in resolution) generates 10 -- 30\% variations in some remnant properties, the overall shapes of the profiles in Fig.~\ref{fig:viscnumcomp} are very similar.  In particular, the merger remnant does not look more or less ``similar-mass'' (except, arguably, the temperature curve at the lowest resolution).  Exact values of properties, therefore, will vary depending on resolution (and will vary on similar or larger levels if initial conditions like \azero\ are changed), but the overall picture of the merger and trends should be more robust.



\section{Comparison With Others}
\label{sec:compwithothers}


\subsection{Comparison With \cite{loreig09}}
\label{ssec:compwithloreig}

\begin{figure}
\centering
\includegraphics[angle=0,width=1.0\columnwidth]{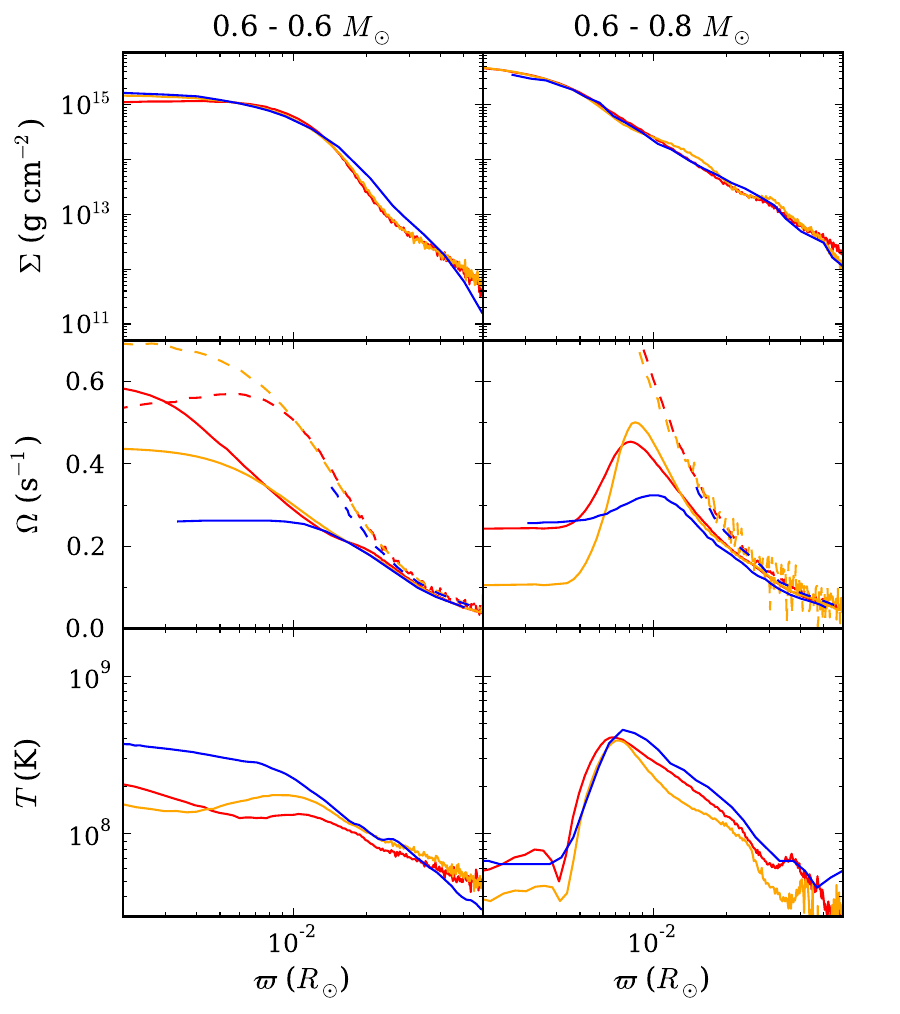}
\caption{Comparison of our results with those of \citetalias{loreig09}, for a 0.6 - 0.6\,\Msun\ (left) and a 0.6 - 0.8\,\Msun\ (right) merger.  Shown are surface density, remnant (solid) and Keplerian (dashed) angular frequency, and temperature, with profiles from \citetalias{loreig09} in blue, and our equivalent ones in red and orange.  Here, the former are for our default completion time of 6 initial orbital periods and the latter for their completion times (514 s or 10.9 orbital periods for the 0.6 - 0.6\,\Msun\ merger, and 164 s or 3.4 orbits for the 0.6 - 0.8\,\Msun\ merger).}
\label{fig:compwithloreig}
\end{figure}

\citeal{loreig09} simulated a number of WD mergers, and gave detailed temperature, surface density, and rotational frequency curves for three.   In Fig.~\ref{fig:compwithloreig}, we compare their results (from their Figs. 3 and 4) with ours for two of these, 0.6 - 0.6\,\Msun\ and 0.6 - 0.8\,\Msun\ (the third was a 0.4 - 0.8\,\Msun\ He - CO WD merger, whose temperature profile cannot be compared directly).  We note that they used different initial conditions, starting their systems with an orbital separation too large for mass transfer to begin, and then slowly reducing the separation until it does.  This point defines their $t = 0$ and the start of the merger simulation proper.  Given this different setup, their merger completion times cannot be compared directly to ours.  In their simulations, however, coalescence (the final consolidation of the two WDs into one) also occurs after just about one orbit, so the differences should not be too large.  To give a sense of the effect of different completion criteria, we compare their results both with our standard results, taken after 6 orbits, and our results taken at their merger completion times.

For both mergers, the surface density curves are similar, although in their 0.6 - 0.6\,\Msun\ merger, the central peak is $\sim\!30$\% higher ($\sim\!10$\% if we use their completion time of 514\,s).  For the 0.6 - 0.8\,\Msun\ merger, the temperature profiles are also very similar, with maxima\footnote{Maximum temperatures given in Table 1 of \citeal{loreig09} refer to hot spots in their simulations, and are about a factor of 2 higher than the hottest points on their temperature curves.  As we have not done hot-spot finding, we cannot compare with those values.} differing by only $\sim$10 - 15\%, and having nearly identical shapes.  Larger differences are seen for the 0.6 - 0.8\,\Msun\ rotational frequency profile, where the angular velocity peaks further out and at lower value ($\sim\!0.3\,\psec$ compared to our $0.45\psec$ -- or $0.50\,\psec$ using their completion time of 164 s).  Indeed, our entire remnant is more spun-up than theirs.  

For the 0.6 - 0.6\,\Msun\ merger, \citeal{loreig09} have a plateau in their angular frequency profile, with $\Omegamax\simeq0.25\,\psec$, while our profile is much more peaked and reaches a much higher frequency, of $0.60\,\psec$.  By their completion time, our rotation curve is not as sharply peaked, but still reaches $0.44\,\psec$.  The temperature profiles are also much less similar: our maximum temperature in the equatorial plane is a factor of 2 lower than theirs (factor of 3.3 at their completion time), and even our maximum temperature along the rotational axis is a factor 1.6 lower (factor 1.9 at their completion time).  

Finally, we can compare how mass is distributed.  In both our simulations and those of \citeal{loreig09}, negligible mass is lost, so only the distribution between disk and core-envelope matters.  For our 0.4 - 0.8\,\Msun, 0.6 - 0.6\,\Msun, and 0.6 - 0.8\,\Msun\ simulations, we infer disk masses  of 0.31, 0.10, and 0.40\,\Msun, respectively, which are reasonably close to the 0.28, 0.10, and 0.30\,\Msun, respectively, listed by \citeal{loreig09} (their Table~1), especially considering that we likely use a different definition of what is ``disk''.

Overall, the primary differences between our simulations appear to be the amount of spin-up and heating of the equal-mass merger.  We believe it is unlikely that this reflects differences in initial conditions: we found much smaller changes in the angular velocity profile with increasing {\azero} (see Fig.~\ref{fig:distcomp}), and in the simulation of \citeal{loreig09} the stars still seem to be quite close to spherically symmetric at the start and disrupt quickly (their Fig.~1), even though they were more properly relaxed.  Instead, we believe the more likely explanation is that the viscosity prescription of \citeal{loreig09}, based on Riemann solvers, yields larger effective viscosity.  This would explain both the reduction in angular velocity and increase in temperature (since viscous evolution converts rotational into thermal energy), as well as the fact that similar-mass mergers are affected more (they mix more, and \citeal{loreig09} ran their equal-mass merger for a very long time).  If we ran our simulations longer and thus included further viscous evolution, the similarity with their simulations would likely be closer.

\subsection{Comparison with Others}
\label{ssec:compwithothers}

Simulations of WD mergers have also been presented by \citet{yoonpr07}, \citet{pakm+10,pakm+11,pakm+12}, \citet{dan+11,dan+12}, and \citet{rask+12}.  Unfortunately, comparison with those results is difficult, since, unlike \citeal{loreig09}, all these authors are sparse with quantitative details about their results.


An exception is the 0.81 - 0.9 {\Msun} merger simulated by \cite{dan+12}, shown in their Fig.~1.  While that simulation is for synchronized WDs, it is still particularly useful to compare with, since \citeauthor{dan+12} show results for both approximate and accurate initial conditions.  We find that their spherically enclosed mass profile is very similar to ours, with, e.g., $M=0.9\,\Msun$ at $4.5\times10^8\,$cm in both (though since spherically enclosed mass is a cumulative quantity, significant structural differences can remain hidden).  Our spherically averaged density profile looks most similar to the profile they found using approximate initial conditions.  Our central density, $1.9\times10^7\,\gcc$, is within $\sim\!10$\% of theirs, and the density profile remains similar up to $r \simeq 5 \times 10^8\,$cm ($\rho\simeq10^6\,\gcc$).  Beyond, their profile becomes shallower while ours continues to decline; at $r=10^9\,$cm, they find $\rho\simeq3\times10^5\,\gcc$, while we find $\rho\simeq10^5\,\gcc$.  This may be a consequence of the additional angular momentum associated with synchronized rotation.  With accurate initial conditions, a difference with our results is that the density profile becomes flat beyond $10^9\,$cm.  

Comparing temperature profiles, we roughly reproduce their spherically averaged one for approximate initial conditions, including the off-center peak -- their \Tmax\ is $\sim\!20\%$ lower (to be expected since their binary is synchronized; see Sec.~\ref{ssec:synchronization}), but is also located at $4\times10^8\,$cm (or $M_r\simeq0.9\,\Msun$).  However, our central temperature ($2.2\times10^8$ K) is an order of magnitude higher than theirs ($2\times10^7$ K) and at $r\gtrsim10^9\,$cm our temperatures are systematically hotter, perhaps a result of the much larger dissipation expected for non-rotating WDs.  With accurate initial conditions they found an even narrower temperature peak than the one with approximate conditions, which thus deviates even more from our curve.  This trend is similar to what we see when increasing \azero\ (Sec.~\ref{ssec:varyingazero}), so it seems likely we would reproduce their simulations more closely if we used the same initial conditions.

\subsection{The Importance of Accurate Initial Conditions}

Many of the recent simulations \citep{dan+11,dan+12,rask+12} assume co-rotating WDs.  This assumption is numerically convenient, in that it is relatively straightforward to start the simulation in the physically correct state: since in the co-rotating frame there are no flow velocities, one can easily relax a simulated binary within an appropriate potential in the co-rotating frame, damping out any velocities resulting from an initial mismatch.  

As a result, it has been possible to study the onset of mass transfer in detail.  As first pointed out by \citet{dsou+06} from simulations using a grid code, the disruption of the donor is preceded by a rather long -- dozens of orbits -- phase of mass transfer.  Further simulations by \cite{dan+11,dan+12} showed that in this initial phase a significant fraction, $\sim10\%$ of the donor mass, is transferred.  As a result, e.g., the disk is substantially colder and more extended.  The remnant core seems more subtly affected, in that its appearance becomes ``more dissimilar'', reflecting that coalescence is between two WDs whose masses have become more disparate than they were initially.  As a consequence, e.g., even for similar-mass binaries, the hottest point of the merger is found to be well outside the center.  Indeed, \citet{rask+12} find that even for equal-mass binaries, the final outcome for more massive mergers is one where the core of one of the WDs is virtually undisturbed.

At present, it is not clear how important accurate initial conditions would be for asynchronous mergers.  Qualitatively, we expect the effects to be smaller than for synchronous mergers, for three reasons.  First, from the analytic study of \cite{lairs94}, in which tidal and rotational distortion are approximated by ellipsoids, co-rotating binaries always reach contact or Roche lobe overflow before becoming dynamically unstable, while irrotational binaries become dynamically unstable first.  While an exact treatment of the irrotational case found that, in fact, Roche contact preceded dynamical instability \citep{uryue98}, it suggests that WDs in irrotational binaries will disrupt much sooner.  Second, the simulations of \citeal{loreig09} use initial conditions that should be quite close to correct, yet their WDs disrupt quickly (see Sec. \ref{ssec:compwithloreig}).  Third, the two components are counterrotating in the rotating frame.  Hence, any mass transferred will hit the accretor with a larger relative velocity than would be the case for co-rotating WDs.  Indeed, in the limit of equal-mass WDs, very little would happen for co-rotating WDs when one reaches contact, while a strong shock would be expected for the irrotational case.  In general, one expects part of the shocked material to enter a high-entropy halo around the accretor.  For co-rotating WDs, \cite{dan+11} found that this halo helps remove angular momentum from the orbit, leading to a shorter start-up phase.  For the irrotational case, given the stronger expected shocks, the start-up phase would likely be reduced even further.  On the other hand, we saw in Sec. \ref{ssec:varyingazero} that remnant properties are sensitive to changes in angular momentum content through changes in \azero.  Simulating more realistically the onset of mass transfer through accurate initial conditions will likely change \azero.

Ideally, one would still simulate the initial mass transfer phase accurately.  Unfortunately, even though the equilibrium solution is known \citep{uryue98}, it is not straightforward to set up the initial binary properly, since it is difficult to relax to a state that includes substantial fluid motion, and to slowly evolve such a state to contact, while ensuring viscosity remains low enough that there is no artificial tidal dissipation.  Such dissipation is seen in our tests with varying initial distance {\azero} in Sec.~\ref{ssec:varyingazero} (and may affect the simulations of \citeal{loreig09} as well).  Prior to coalescence, strong dissipation of tidal bulges heats the outer envelope of the donor (both stars for similar-mass mergers), and spin-orbit coupling due to both tides and the direct-impact accretion stream result in both donor and accretor becoming 25 - 50\% synchronized by coalescence.

Since it significantly affects the merger and merger outcome, whether or not tidal dissipation causes real CO WD binaries to synchronize before the merger remains a major source of uncertainty.  For the radiative stellar envelopes appropriate for WDs, tidal dissipation is expected to be inefficient, with a timescale $10^{12}$ to $10^{15}$ yrs, suggesting that WDs do not synchronize (\citealt{marsns04}, and references therein).  However, coupling of the tides to pulsations may dramatically increase dissipation \citep{fulll12}.  Fortunately, it may be possible to determine the rate of synchronization observationally.  For instance, \citet{piro11} suggested that tidal dissipation is responsible for the relatively high temperature of the primary WD in the 13-minute eclipsing binary SDSS J065133.33+284423.3, predicting that it would be about halfway to being synchronized.  This could be tested by either measuring the rotational broadening of the narrow cores of the hydrogen lines, or looking for velocity deviations through the transit of the more massive secondary.




\section{Post-Merger Evolution}
\label{sec:postmerger}

\begin{figure*}
\centering
\includegraphics[angle=0,width=2.0\columnwidth]{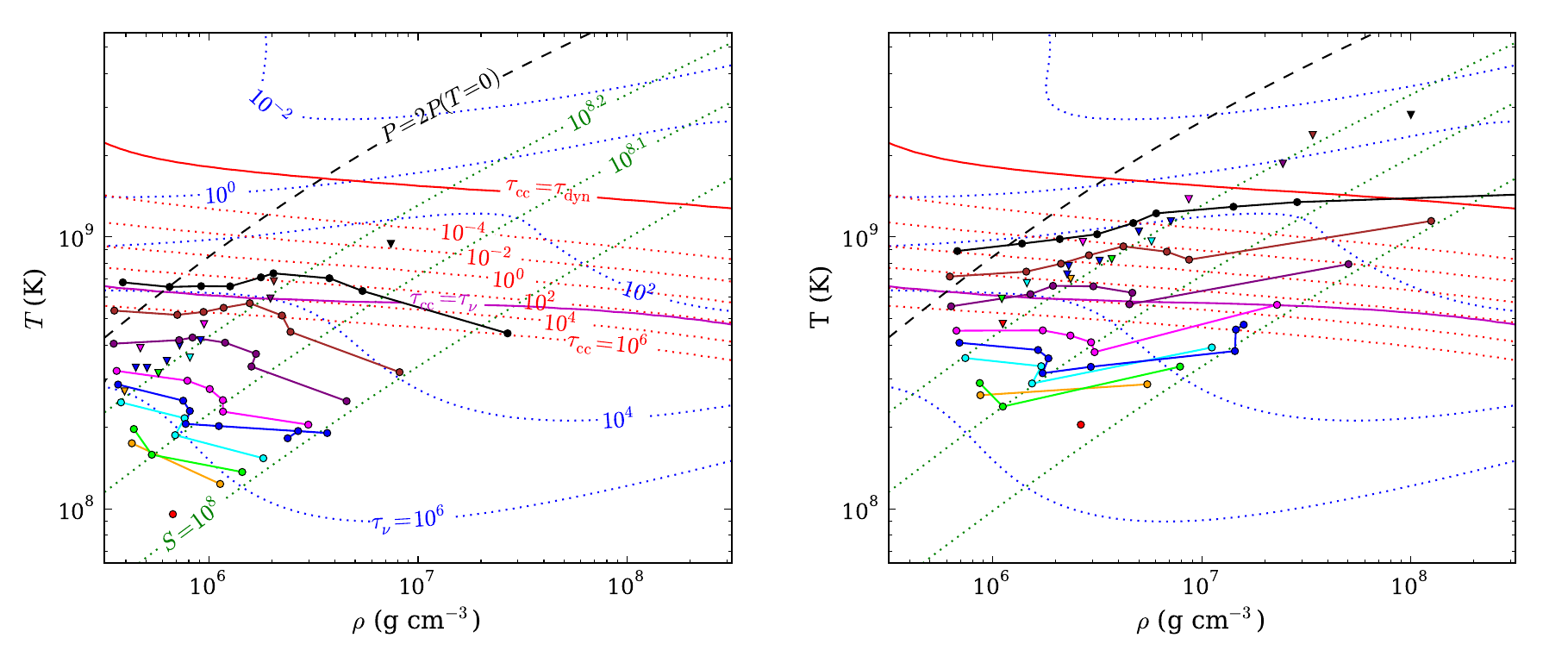}
\caption{Left: merger remnant maximum temperature {\Tmax} and corresponding density {\rhoTmax} for all merger remnants.  Values along the equatorial plane are marked with circles, with lines connecting points with the same accretor mass, while values along the rotational axis (only plotted for similar-mass mergers) are marked with triangles (for all, colors indicate accretor mass, encoded as in Fig.~\ref{fig:constacc}).  For similar-mass mergers, equatorial temperatures have been adjusted to account for mixing in convectively unstable cores.  Right: maximum temperatures and corresponding densities following estimated post-merger evolution.  The estimate assumes that the remnant spins down completely, that all rotational energy is used to drive matter to large distances, and that the remainder adjusts adiabatically (see text).  Also shown are contours of constant neutrino cooling timescale $\tau_\mathrm{\nu}\equiv C_P T/\varepsilon_\nu$ and carbon fusion heating timescale $\tau_\mathrm{cc}\equiv C_P T/\varepsilon_{\rm CC}$, both in years, as well as entropy $S$ in ${\rm erg\,K^{-1}}$.  (Here, $C_P$ is the heat capacity at constant pressure and $\epsilon$ the specific energy loss/gain rate.)  The lines labeled $\tau_\mathrm{cc} = \tau_\mathrm{\nu}$ and $\tau_\mathrm{cc} = \tau_\mathrm{dyn}$ denote where the carbon fusion heating timescale balances the neutrino cooling and dynamical timescales, respectively.  Finally, the $P = 2P(T$=$0)$ line is shown as an approximate upper bound of the region where degeneracy pressure dominates.  All quantities were calculated using MESA \citep{paxt+11}.}
\label{fig:willitexplode}
\end{figure*}



\begin{figure}
\centering
\includegraphics[angle=0,width=1.0\columnwidth]{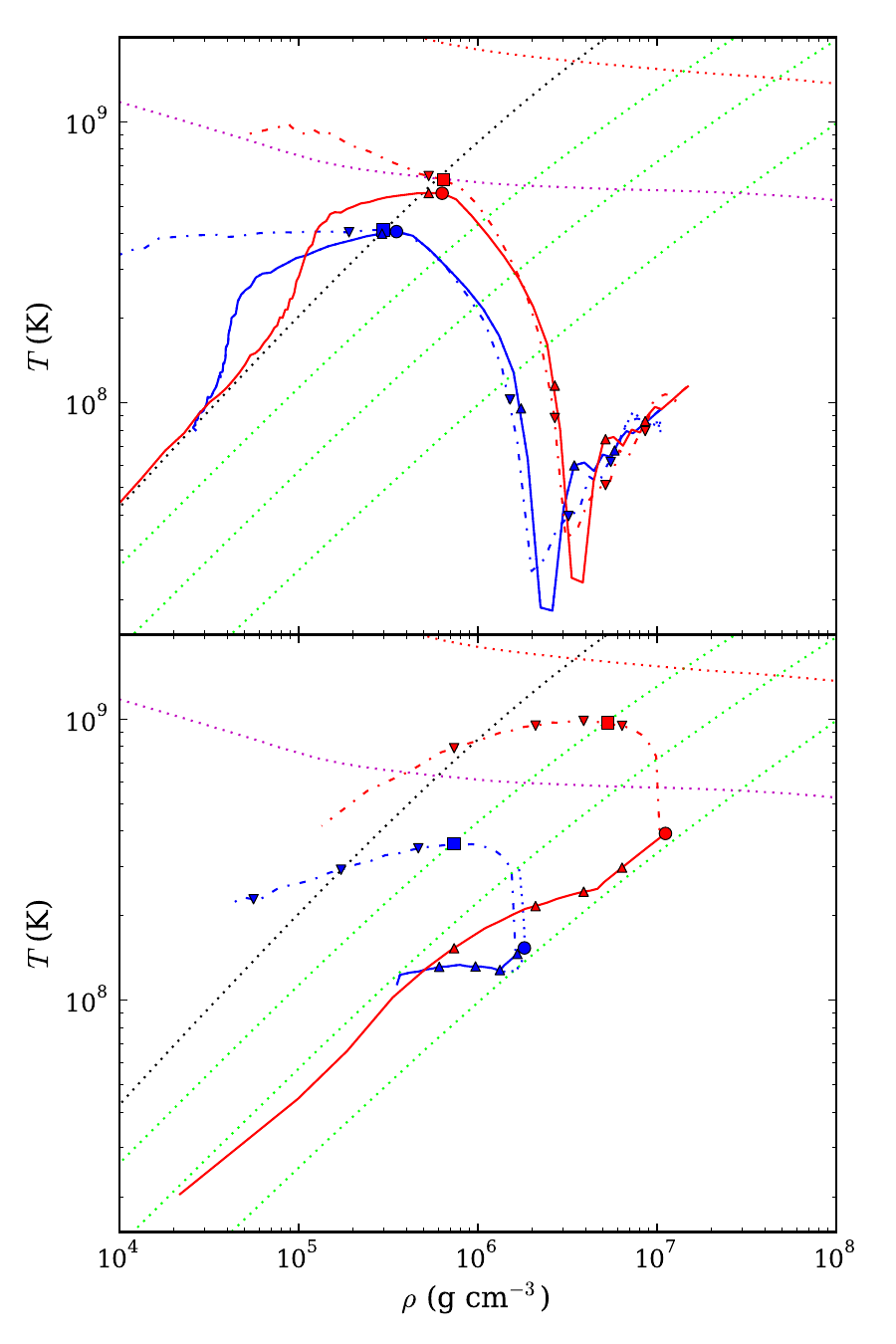}
\caption{Estimate of post-merger viscous evolution for 0.4 - 0.8 {\Msun} (top) and 0.6 - 0.6 {\Msun} (bottom) mergers.  In blue are shown the temperature-density structure of the merger remnant, on the equatorial plane before (dotted) and after (solid) correction for convection, as well as along the rotational axis (dot-dashed), with points marking the hottest locations (circles and squares) and steps of 0.2\,\Msun\ in spherical enclosed mass (triangles pointing up and down).  In red, estimates of the structure following viscous evolution are shown, where it is assumed that the remnant spins down completely, that all rotational energy is used to drive matter to large distances, and that the remainder adjusts adiabatically (see text).  For reference, also shown as dotted curves are the contours of constant entropy from Fig.~\ref{fig:willitexplode} (green), as well as the lines where $\tau_\mathrm{cc} = \tau_\mathrm{\nu}$ (magenta), $\tau_\mathrm{cc} = \tau_\mathrm{dyn}$ (red), and $P = 2P(T$=$0)$ (black).}
\label{fig:pmevolution}
\end{figure}

We now turn to the question of how our merger remnants will evolve.  To set the stage, we show in the left panel of Fig.~\ref{fig:willitexplode} for all remnants the maximum temperature \Tmax\ found along the equatorial plane\footnote{The central temperatures for the 0.625 - 0.65 \Msun\ and 1.0 - 1.0 \Msun\ mergers are $\sim\!4$\% and 10\% lower than their respective maximum temperatures.  In both cases, however, the center is much denser than the off-center hotspot, and since our estimated post-merger evolution more greatly affects central material, we show the central equatorial density and temperature for these two systems in Fig. \ref{fig:willitexplode} left, rather than the maximum.} as a function of the corresponding density \rhoTmax.  Here, for the similar-mass mergers for which we found convectively unstable cores (Sec.~\ref{sssec:thermtrends}), we show the (lower) temperatures reached after artificially mixing them.   For those mergers, the much higher temperatures reached along the rotational axis are shown as well (triangles).  One sees the trends identified earlier: \Tmax\ is mostly set by the accretor, while \rhoTmax\ depends more strongly on the donor.  As a result, maximum temperature occurs in less degenerate conditions for dissimilar-mass mergers, crossing the degeneracy line for our most disparate cases.  One also sees that for all but the most massive accretors, carbon fusion will not start: the neutrino cooling time is shorter than the fusion heating time.  This is consistent with what was found in previous work (see Sec.~\ref{sec:intro}).

\subsection{Viscous Evolution and Possible Spin Down}

Following the merger, processes that happen on timescales slower than the dynamical time can become important.  These include viscous evolution, neutrino emission, radiative or convective thermal adjustment, and magnetic dipole radiation spin-down.  Out of these, convection acts on the fastest timescale, and we already included its effect on the core in Fig.~\ref{fig:willitexplode} left.  Next fastest would almost certainly be viscous evolution.  The merger remnant is unstable to both the magneto-rotational instability \citep{balbh91} and Tayler-Spruit dynamo \citep{spru02}.  Radiative adjustment is expected to be much slower, except at the surface, where radiative losses may also lead to convection in some systems \citep{shen+12,schw+12,rask+12}.  Using the standard \cite{shaks73} $\alpha$-prescription for the viscosity $\nu = \alpha c_s H$, where $c_s$ is the local soundspeed and $H$ is the scale height of the system, the viscous evolution timescale for the remnant disk is
\begin{equation}
t_\mathrm{visc} = \frac{R_\mathrm{disk}^2}{\nu} = \frac{1}{\alpha}\left(\frac{R_\mathrm{disk}}{H}\right)^2t_\mathrm{dyn} \sim \frac{10}{\alpha}t_\mathrm{dyn},
\end{equation}
implying a timescale $t_{\rm visc}\sim10^3-10^5{\rm\,s}$ for $\alpha\sim10^{-3}-10^{-1}$ and $t_\mrm{dyn} \sim 10$ s.  This is orders of magnitude smaller than both the neutrino loss timescale ($\gtrsim\!10^3$ yrs; see Fig.~\ref{fig:willitexplode} left) and thermal adjustment timescale ($\gtrsim\!10^4$ yrs; \citealt{shen+12}).

It is possible that the strong differential rotation during a merger results in substantial amplification of magnetic fields.  The one known probable WD merger remnant, RE J0317$-$853, has a surface magnetic field of $3.4\times10^8\,$G \citep{bars+95,kube+10}.  If mergers lead to strongly magnetized WDs, and these WDs additionally drive an ionized outflow, the magnetic coupling between the outflow and the WD could serve to transport angular momentum out of the system, spinning down the WD.  The timescale for such a spin-down is roughly given by,
\begin{equation}
t_\mathrm{msd} \sim \frac{L}{\dot{M}R_A^2\Omega}
              \sim \frac{L}{(\dot M\Omega)^{3/5}(B^2R^6)^{2/5}},
\end{equation}
where $L$ is the angular momentum of the remnant, $\dot M$ the mass loss rate, $R_A$ the Alfv\'{e}n radius, and $\Omega$ the angular spin frequency.  For the second approximation, we used that $R_A\sim (B^2R^6/\dot M\Omega)^{1/5}$, with
$B$ the surface magnetic field and $R$ the remnant radius.  Scaling to  $B=B_810^8\,$G and $\dot M={\dot M}_{-7}10^{-7}\,\Msun{\rm\,yr^{-1}}$ (similar to what is observed for RE J0317$-$853 -- see above -- and [WR] cores of planetary nebulae [\citealt{hama97}]), and using the properties inferred for a 0.6 - 0.6 \Msun\ remnant ($L\simeq L_{\rm tot}\simeq10^{50.5}{\rm\,g\,cm^2\,s^{-1}}$, $R\simeq R_{\rm disk}\simeq10^9{\rm\,cm}$, $\Omega\simeq\Omega_{\rm max}\simeq10^{-0.3}{\rm\,s^{-1}}$, we find $t_{\rm msd}\simeq8\times10^3\,B_8^{-4/5}{\dot M}_{-7}^{-3/5}{\rm\,yr}$, which is of the same order as the neutrino cooling timescale of $\sim\!10^4{\rm\,yr}$ at the ignition line (for the whole range of remnants, $2\times10^3\lesssim t_{\rm msd}\lesssim 5\times10^4{\rm\,yr}$).

Accretion from the disk, loss of rotational support, and possible cooling of the hot envelope could all compress and heat the remnant core.  A detailed study of this is beyond the scope of this paper, but we can make first-order estimates of the effects on our merger remnants, and compare these with the more detailed analysis of \cite{schw+12} in one specific case.

For our estimates, we make four assumptions: (i) spin-down and accretion are much faster than thermal processes, and do not lead to local dissipation (i.e., particles entropies are constant in time); (ii) all angular momentum is carried away to large distances; and (iii) corresponding matter ends up with zero total energy (i.e., is at large distances and has negligible kinetic and internal energy).  From energy conservation, the last assumption implies that the remaining object will have the same total energy as our merger remnant (but a lower mass), the first that it will have the same entropy structure, and the second that it has no rotational support.  To determine the properties, we first determine the entropy profile of the merger remnant, by averaging entropy over isopotential surfaces.  We then use this entropy profile and an estimated central density to construct a spherically symmetric (non-spinning) hydrostatic model, iterating on the central density until it has the correct total energy (inside of the zero-pressure surface).  This automatically gives the mass contained in this object, which will be lower than our remnant mass, the remainder representing material that, due to dissipation of rotational energy, has expanded out to large distances and therefore provides negligible weight.  To determine the evolution of hot spots, we order remnant particles by potential, and map them to their new positions in the final object, calculating new temperatures from the new densities, again assuming their entropy did not change (entropy is not constant over isopotential surfaces, so these temperatures are not strictly consistent with the hydrostatic model).

In Fig.~\ref{fig:pmevolution}, we show the results of our evolutionary estimate for our fiducial 0.4 - 0.8 and 0.6 - 0.6 {\Msun} systems.  For the former case, the core-envelope, originally 0.90\,\Msun, accretes 0.06\,\Msun\ from the disk, the remaining 0.25\,\Msun\ going to large distances.  The central core is not significantly heated, while the lower-density hot envelope is, with the outer hot envelope along the rotational axis passing the ignition line.  Since this material is almost non-degenerate, the resulting nuclear burning will likely be stable, or be extinguished by expansion.  Thus, not unexpectedly, the hot envelopes of dissimilar-mass mergers are not good candidates for a nuclear runaway.  


For the 0.6 - 0.6 {\Msun} system, the center of the final, spun-down object is at much higher density and temperature than the remnant, while much of the outer regions have become less dense and cool.  The latter happens because similar-mass mergers have strong rotational support, and if this is removed their binding energy increases significantly.  To compensate for this, a large amount of mass has to expand to large distances, causing the core-envelope mass to decrease from 1.11 {\Msun} to 0.91 {\Msun}.  In the final object, the hottest point on the equatorial plane does not reach the ignition line, but the significantly hotter points above and below the equatorial plane do, at densities under which degeneracy pressure still dominates.  Hence, if the hot spots indeed compress with the rest of the remnant, a nuclear runaway could be triggered.  (Of course, a nuclear runaway would start as soon as the heating timescale becomes shorter than the compression timescale, which may happen closer to the ignition line.)


In the right panel of Fig.~\ref{fig:willitexplode}, we show the results of applying our estimates to all our merger remnants.  One sees that all compress and heat, and almost every remnant whose accretor mass is above 0.8\,\Msun\ will reach ignition somewhere on the equatorial plane, in many cases under degenerate conditions.  We also chart the evolution of the off-center hot spots in similar-mass mergers (square points), and while they are at lower density, they remain degenerate and are all pushed substantially further above the ignition line than their counterparts on the equatorial plane.  Almost all similar-mass mergers with an accretor mass above 0.5 {\Msun} could therefore experience nuclear runaways due to their hot spots, though at least some of them will become non-degenerate before an explosion can occur.

The above suggests it is at least plausible that many of our mergers would eventually ignite in degenerate conditions, and that it thus is worthwhile to simulate their evolution in detail.  Suitable simulations have recently been pioneered by \cite{shen+12} and \cite{schw+12}.  \citeauthor{shen+12} started with a one-dimensional simulation, where they ported the remnant of a 0.6 - 0.9 {\Msun} merger (from \citealt{dan+11}), and evolved it assuming a $\gamma = 5/3$ polytropic equation of state and an $\alpha = 10^{-2}$ viscosity.  They find the system spins down completely due to outward angular momentum transport, and the rotationally-supported thick disk is transformed into a tenuous, thermally-supported envelope that hardly affects the core.  Over longer, thermal evolution timescales (simulated using MESA, \citealt{paxt+11}), this tenuous hot envelope cools, compresses the core, and lights off-center convective carbon burning, eventually turning the remnant into an ONe WD (that may end its life in an accretion induced collapse).

\cite{schw+12} went a step further, porting the same 0.6 - 0.9 {\Msun} simulation, as well as seven other systems, into two-dimensional ZEUS-MP2 simulations \citep{haye+06}, using the Helmholtz equation of state and an $\alpha=3\times10^{-2}$ viscosity.  They confirm the one-dimensional results, finding complete spin-down and transformation of the rotationally supported disk into a tenuous, spherically symmetric, hot envelope.  They find a 50\% increase in the temperature of the hottest point, and a factor of 3 increase in the corresponding density.  They also find entropy to roughly be constant in the remnant, except in the outer regions and at the very center, where dissipation of rotational energy leads to heating.

It is encouraging that the results of the above detailed simulations are similar to what we find using our first-order estimates.  For our 0.6 - 0.9 {\Msun} remnant, our estimate give increases for the hottest equatorial point of a factor of 2.5 in density and 1.6 in temperature, reasonably close to what is found by \citeauthor{schw+12}\footnote{For the hottest point along the rotational axis, we find a factor of 4.8 increase in density and 2.1 increase in temperature, suggesting our model does not depict as well the evolution of the outer hot envelope along the rotational axis.}.  Thus, our simplifying assumptions appear to be appropriate at least for dissimilar-mass mergers, where most of the rotational dissipation will be in the disk and envelope, and the structure of the remainder is roughly spherically symmetric (both in density and temperature).  It is not clear our estimates would be equally good for similar-mass mergers, where rotational dissipation should occur throughout the star, heating the entire remnant, and where there are substantial differences between the remnant's density and temperature structures.  In particular, the rotational axis hotspots may dissipate, which would potentially make it more difficult for a similar-mass system to achieve a runaway.  It will thus be particularly interesting to simulate the further evolution of those remnants in more detail.  Unfortunately, no such remnants were included by \citeauthor{shen+12} and \citeauthor{schw+12}.

\subsection{Possible Explosions?}

From our estimates, it seems that, as suggested by \citeal{vkercj10}, many merger remnants will ignite carbon fusion.  If a detonation is triggered, the resulting explosion may well resemble an SN Ia.  Indeed, if the remnants spun down before ignition, their structures are sufficiently close to that of a cold WD that the calculations of \cite{sim+10} should apply.  From our estimates, for mergers that have a total mass between 1.2 and 1.4\,\Msun\ (which should be the most common ones), the final objects have masses between $\sim\!0.9$ and $\sim\!1.1$ \Msun\, which matches fairly nicely the range of $\sim\!1$ to $\sim\!1.2$ \Msun\ required to reproduce the observed range of SN Ia luminosities \citep{sim+10}.


Of course, it is far from clear whether ignition leads to a detonation, since we do not currently understand how detonations are triggered (\citealt{seit+09,woos+11}, and references therein).  Generally, it should help that ignition in our remnants is at much lower density (a few $10^7\,{\rm\,g\,cm^{-3}}$) than is the case for near-Chandrasekhar models ($\sim\!10^9{\rm\,g\,cm^{-3}}$), because complete burning leads to much larger relative overpressures (e.g. \citealt{mazumw77}; \citealt{seit+09}).  Also, if a deflagration is started, plausible mechanisms to transition to a detonation all seem to require densities around $10^7\,{\rm\,g\,cm^{-3}}$, where the conductive flame speeds are slower and the separation between the various burning fronts increases (e.g., \citealt{woos+09,woos+11}).

An interesting aspect of our results is that for all cases ignition likely happens off-center: in shells for dissimilar-mass mergers and in hot spots along the rotational axis for similar-mass ones.  Previous one-dimensional simulations suggested off-center ignition would lead to a slow deflagration flame that turns the CO WD into a ONe WD (e.g., \citealt{saion85}).  However, these calculations assumed a hot spot many pressure scale heights above the center.  For ignition closer to the center, a deflagration plume is produced (e.g., \citealt{aspd+11}), which may transition to a detonation \citep{seitcr11} and unbind the star.   

Given our findings, it seems likely that, if a detonation occurs, it will be triggered off-center.  It would be interesting to simulate the resulting explosion, and see whether one could reproduce the observational evidence for asymmetries, which have been interpreted in terms of off-center ignition (though so far only in the context of near-Chandrasekhar models; \citealt{maed+10a,maed+10b}).

Finally, while we simulated only mergers of CO WDs, we can extrapolate our results to more massive ONe WDs.  For these, the temperatures would be at least as high as for our 1\,\Msun\ accretors, and, after further viscous evolution, the mergers should become hot enough to ignite Ne burning.  If this also leads to a detonation, the lower fusion energy released would likely lead to a less energetic explosion than expected for a CO WD merger, but it would produce far more $^{56}$Ni and have a very large mass.  Plausibly, it would resemble an SN Ia like SN 2009dc, which had unusually low ejecta velocities, produced $\sim\!1.8\,\Msun$ of $^{56}$Ni and had a total ejecta mass of $\sim\!2.8\,\Msun$ \citep{taub+09}.

\section{Conclusion}

We have performed a large, detailed parameter-space study of CO WD mergers, extracting pertinent properties and profiles for each remnant, and studying how these vary across parameter space.  For a merger involving dissimilar-mass WDs, with low {\qrho} = {\rhocrat}, the outcome is a cold, slowly rotating, degeneracy-supported remnant core, which is essentially unaffected by the merger, surrounded by a hot, roughly spherical envelope and, further out, by a sub-Keplerian disk.  For a similar-mass merger, with high {\qrho}, an ellipsoidal core is produced along with a small disk, and the entire remnant is hot and partly supported by rotation.  The transition between these two regimes is smooth, but occurs roughly at $\qrho\simeq0.6$, or equivalently a mass difference $\Delta M= \Ma-\Md \simeq 0.1$.  We found that for a fixed {\qrho}, merger remnant curves are roughly homologous.  We also presented trends for a number of merger remnant properties, providing linear scaling relations and best fits for most of them, hoping these can guide theoretical understanding and help analytical estimates.

We made first-order estimates of the post-merger viscous evolution and spindown, and found that it is plausible that a large fraction of the mergers simulated will eventually experience a nuclear runaway, as was suggested by \citeal{vkercj10}, and thus possibly end as thermonuclear supernovae.  Further, detailed, simulations of this evolution across the whole parameter space, using techniques similar to those of \citet{shen+12} and \citet{schw+12}, would be required to confirm this.  If the evolution of these remnants results in a detonation, a detailed comparison of the resulting light curve with observations must be carried out.

Our work represents one of the most detailed parameter studies of WD mergers to date.  It would benefit, however, from resolution of a number of topics.  First, for greater precision, it will be necessary to use better initial conditions.  For synchronized systems, it is already known this has nontrivial effect on the outcome \citep{dan+11,dan+12}, and our results suggest it is important also for non-synchronized systems.  Unfortunately, for the non-synchronized case, it is not trivial to implement the initial conditions, but better approximations are possible.  Second, it would be useful to try to compare with merger simulations done with a grid code, which should have become more straightforward now that good moving mesh codes have become available \citep{spri10,duffm11}.  More generally, whether or not WDs are synchronized before the merger remains unknown, yet clearly affects the resulting merger.  Hopefully, this can be resolved empirically, by measuring the spin frequency for WDs in the very short-period binaries that have recently been discovered (e.g., \citealt{brow+11}).

We thank Pablo Lor\'{e}n-Aguilar, Enrique Garc\'{i}a-Berro, Stuart Sim, Enrico Ramirez-Ruiz, Marius Dan, James Guillochon, Evan Scannapieco, Cody Raskin and Ken Shen for insight into their simulations and useful discussion on the physics of mergers and post-merger evolution.  We are grateful to Frank Timmes for creating the Helmholtz equation of state, and assisting us with its implementation in Gasoline, as well as to Bill Paxton and the MESA team for creating MESA and making it modular.  This work made extensive use of NASA's ADS and was supported by the Vanier and Discovery grants of Canada's Natural Sciences and Engineering Research Council (NSERC).

\section{Appendix}
\label{sec:appendix}

We present CO WD binary pre-merger properties in Tab. \ref{tab:input1} and \ref{tab:input2}, and post-merger remnant properties in Tab.~\ref{tab:output1} -- \ref{tab:output5}.  \notetoeditor{We would like to merge Tables 1 and 2 into a new, giant Table 1, and Tables 3 -- 7 into a new Table 2.}

Masses and distances are in {\Msun} and {\Rsun} ({\Msun} $=\,1.9891\times10^{33}$ g, {\Rsun} $=\,6.95508 \times 10^{10}$ cm), and all other quantities are in CGS.  Numerical subscripts indicate powers of 10.  

Subscript ``d'' indicates donor, and ``a'' accretor, while sub/superscript ``disk'' and ``ce'' stand for the disk and core-envelope, respectively.  Superscript ``z'' indicates a value along the rotational axis.  Subscript ``max'' stands for the maximum value of a property, either along the equatorial plane if standalone or along the rotational axis if paired with superscript ``z''.  Superscript ``cv'' indicates an estimate of a property that takes into account post-merger convection along the equatorial plane (convection does not occur along the rotational axis).  Subscript ``rot'' stands for rotational, ``th'' thermal, ``deg'' degeneracy, and ``pot'' gravitational potential energy.

The disk inner radius $\varpi_\mrm{disk}$ (Tab.~\ref{tab:output5}) is determined by flattening the system along the $z$-axis, and finding the {\rxy} at which the centripetal force counteracts half the gravitational force on the particles.  Bound particles with {\rxy} larger and smaller than this value are part of the disk and core-envelope, respectively (from this differentiation, disk and core-envelope bulk properties are determined).

All bulk energies ($E$), moments of inertia ($I$) and angular momenta ($L$) are determined by summing up individual particle values.  Scaleheights are determined by fitting a Gaussian to the density profile of particles within a cylindrical shell of radius {\rxy}, except for \hxy\ which uses particles in the equatorial plane.  Enclosed masses ($M_\mrm{enc}$) are determined spherically (e.g. $M_\mrm{enc}(T_\mrm{max})$ is the mass enclosed by a sphere of $r = \varpi(T_\mrm{max})$).  In particular, energy enclosed masses $M_\mrm{enc}(fE)$ are the spherical masses enclosing fraction $f$ of energy $E$ of the remnant (e.g. $M_\mrm{enc}(\frac{1}{2}E_\mrm{th})$ is the spherical mass enclosing half of the total remnant thermal energy).

In Tab.~\ref{tab:output1}, {\rhoc} is determined by averaging values of particles within $r < \frac{1}{10}\hz$ of the center.  This accounts for minor discrepancies between {\rhoc} and $\rho_\mrm{max}$, which is determined along the equatorial plane.  $M_\mrm{ej}$ is the total mass of particles that are unbound, which is determined by finding if their total energy $E = E_\mrm{pot} + E_\mrm{K} \geq 0$.

In Tab.~\ref{tab:output2}, central core mass $M_\mrm{cc}$ represents the spherical mass enclosed within a radius $r = \hz$, while $M_\mrm{cc,d}$ represents the donor mass enclosed; $M_\mrm{cc,d}/(M_\mrm{cc} - M_\mrm{cc,d}) = (f_\mrm{d}/f_\mrm{a})_\mrm{cc}$ (Sec. \ref{sssec:whatisequalmass}).

In Tab.~\ref{tab:output3}, {\Tc} is determined in the same manner as {\rhoc}, and therefore has discrepancies with the equatorially-determined {\Tmax} when maximum temperature is at the center of the remnant.

In Tab.~\ref{tab:output4}, for some similar-mass mergers $\rho(T^\mrm{cv}_\mrm{max})$ will be larger than {\rhoTmax} ($\varpi(T_\mrm{max})$ will also change); this is because after six orbital periods these mergers will have their maximum temperature, but not necessarily their maximum density, exactly at the center of the remnant.  We assume that the region undergoing convection will evolve until it becomes isentropic; as a result maximum temperature after convection will shift to the point of maximum density.

In Tab.~\ref{tab:output5}, $\varpi_\mathrm{md}$ is the cylindrical radius enclosing half the mass of the disk.

The ``0.625 - 0.65 255k(2)'' simulation is a re-running of the parameter space 0.625 - 0.65 \Msun\ run, to determine the extent to which random noise affects remnant properties (Sec. \ref{ssec:restest}).

\begin{deluxetable*}{cccccccccc}
\tablecolumns{10} 
\tablewidth{0pc}
\tabletypesize{\scriptsize}
\tablecaption{Table of WD binary initial conditions.}
\tablehead{Simulation & $M_\mathrm{d}$ & $R_\mathrm{d}$ & $(\rho_\mathrm{c,d})_6$ & $(E_\mathrm{rot,d})_{48}$ & $(E_\mathrm{th,d})_{48}$ & $(E_\mathrm{deg,d})_{48}$ & $(E_\mathrm{pot,d})_{48}$ & $(I_\mathrm{d})_{51}$ & $h_\mathrm{d}$}
\startdata
0.4 - 0.4 & 0.4023 & 0.0138 & 1.14 & 0.00 & 0.09 & 18.37 & -35.92 & 0.17 & 0.0053 \\
0.4 - 0.5 & 0.4023 & 0.0138 & 1.14 & 0.00 & 0.09 & 18.37 & -35.92 & 0.17 & 0.0053 \\
0.5 - 0.5 & 0.5026 & 0.0124 & 2.08 & 0.00 & 0.12 & 33.32 & -63.23 & 0.17 & 0.0046 \\
0.4 - 0.55 & 0.4023 & 0.0138 & 1.14 & 0.00 & 0.09 & 18.37 & -35.92 & 0.17 & 0.0053 \\
0.5 - 0.55 & 0.5026 & 0.0124 & 2.08 & 0.00 & 0.12 & 33.32 & -63.23 & 0.17 & 0.0046 \\
0.55 - 0.55 & 0.5533 & 0.0118 & 2.80 & 0.00 & 0.14 & 43.53 & -81.29 & 0.16 & 0.0043 \\
0.4 - 0.6 & 0.4023 & 0.0138 & 1.14 & 0.00 & 0.09 & 18.37 & -35.92 & 0.17 & 0.0053 \\
0.5 - 0.6 & 0.5026 & 0.0124 & 2.08 & 0.00 & 0.12 & 33.32 & -63.23 & 0.17 & 0.0046 \\
\enddata 
\label{tab:input1}
\end{deluxetable*}

\begin{deluxetable*}{cccccccccccc}
\tablecolumns{12} 
\tablewidth{0pc}
\tabletypesize{\scriptsize}
\tablecaption{Table of WD binary initial conditions, continued.}
\tablehead{Simulation & $M_\mathrm{a}$ & $R_\mathrm{a}$ & $(\rho_\mathrm{c,a})_6$ & $(E_\mathrm{rot,a})_{48}$ & $(E_\mathrm{th,a})_{48}$ & $(E_\mathrm{deg,a})_{48}$ & $(E_\mathrm{pot,a})_{48}$ & $(I_\mathrm{a})_{51}$ & $h_\mathrm{a}$ & $a_0$ & $P_\mathrm{orb}$}
\startdata
0.4 - 0.4 & 0.4023 & 0.0138 & 1.14 & 0.00 & 0.09 & 18.37 & -35.92 & 0.17 & 0.0053 & 0.0364 & 77.57 \\
0.4 - 0.5 & 0.5026 & 0.0124 & 2.08 & 0.00 & 0.12 & 33.32 & -63.23 & 0.17 & 0.0046 & 0.0384 & 79.05 \\
0.5 - 0.5 & 0.5026 & 0.0124 & 2.08 & 0.00 & 0.12 & 33.32 & -63.23 & 0.17 & 0.0046 & 0.0328 & 59.14 \\
0.4 - 0.55 & 0.5533 & 0.0118 & 2.80 & 0.00 & 0.14 & 43.53 & -81.29 & 0.16 & 0.0043 & 0.0393 & 79.61 \\
0.5 - 0.55 & 0.5533 & 0.0118 & 2.80 & 0.00 & 0.14 & 43.53 & -81.29 & 0.16 & 0.0043 & 0.0335 & 59.65 \\
0.55 - 0.55 & 0.5533 & 0.0118 & 2.80 & 0.00 & 0.14 & 43.53 & -81.29 & 0.16 & 0.0043 & 0.0312 & 52.42 \\
0.4 - 0.6 & 0.6035 & 0.0114 & 3.67 & 0.00 & 0.16 & 55.72 & -102.47 & 0.16 & 0.0040 & 0.0401 & 80.07 \\
0.5 - 0.6 & 0.6035 & 0.0114 & 3.67 & 0.00 & 0.16 & 55.72 & -102.47 & 0.16 & 0.0040 & 0.0342 & 60.08 \\
\enddata
\label{tab:input2}
\end{deluxetable*}

{\setlength{\tabcolsep}{4pt}
\begin{deluxetable*}{ccccccccccccc}
\tablecolumns{13} 
\tablewidth{0pc}
\tabletypesize{\scriptsize}
\tablecaption{Table of merger remnant properties.}
\tablehead{Simulation & Time & $(\rho_\mathrm{c})_6$ & $(\rho_\mathrm{max})_6$ & $h_\mathrm{z}$ & $h_\varpi$ & $M_\mathrm{ce}$ & $M_\mathrm{ej}$ & $M_\mathrm{disk}$ & $M_\mathrm{disk,d}$ & $M_\mathrm{enc}(\frac{1}{4}M_\mathrm{d})$ & $M_\mathrm{enc}(\frac{1}{2}M_\mathrm{d})$ & $M_\mathrm{enc}(\frac{3}{4}M_\mathrm{d})$}
\startdata
0.4 - 0.4 & 466.3 & 0.69 & 0.67 & 0.0044 & 0.0092 & 0.733 & 0.000 & 0.071 & 0.036 & 0.200 & 0.403 & 0.605 \\
0.4 - 0.5 & 474.4 & 2.13 & 2.05 & 0.0042 & 0.0052 & 0.700 & 0.000 & 0.205 & 0.168 & 0.512 & 0.655 & 0.790 \\
0.5 - 0.5 & 354.3 & 1.13 & 1.13 & 0.0039 & 0.0085 & 0.923 & 0.001 & 0.081 & 0.044 & 0.251 & 0.506 & 0.762 \\
0.4 - 0.55 & 478.5 & 2.87 & 2.76 & 0.0040 & 0.0046 & 0.726 & 0.001 & 0.229 & 0.192 & 0.572 & 0.714 & 0.845 \\
0.5 - 0.55 & 358.4 & 2.37 & 2.29 & 0.0038 & 0.0059 & 0.879 & 0.002 & 0.175 & 0.144 & 0.537 & 0.733 & 0.904 \\
0.55 - 0.55 & 313.6 & 1.44 & 1.44 & 0.0036 & 0.0081 & 1.018 & 0.001 & 0.087 & 0.043 & 0.280 & 0.555 & 0.832 \\
0.4 - 0.6 & 480.5 & 3.94 & 3.78 & 0.0038 & 0.0041 & 0.719 & 0.001 & 0.286 & 0.251 & 0.659 & 0.777 & 0.898 \\
0.5 - 0.6 & 360.4 & 3.59 & 3.45 & 0.0037 & 0.0047 & 0.876 & 0.001 & 0.229 & 0.185 & 0.608 & 0.787 & 0.960 \\
\enddata
\label{tab:output1}
\end{deluxetable*}
}

{\setlength{\tabcolsep}{4pt}
\begin{deluxetable*}{ccccccccccccc}
\tablecolumns{13}
\tablewidth{0pc}
\tabletypesize{\scriptsize}
\tablecaption{Table of merger remnant properties, continued.}
\tablehead{Simulation & $M_\mathrm{cc}$ & $M_\mathrm{cc,d}$ & $(E_\mathrm{rot})_{48}$ & $(E_\mathrm{th})_{48}$ & $(E_\mathrm{deg})_{48}$ & $(E_\mathrm{pot})_{48}$ & $(E^\mathrm{disk}_\mathrm{rot})_{48}$ & $(E^\mathrm{disk}_\mathrm{th})_{48}$ & $(E^\mathrm{disk}_\mathrm{deg})_{48}$ & $(E^\mathrm{ce}_\mathrm{rot})_{48}$ & $(E^\mathrm{ce}_\mathrm{th})_{48}$ & $(E^\mathrm{ce}_\mathrm{deg})_{48}$}
\startdata
0.4 - 0.4 & 0.038 & 0.019 & 14.58 & 4.56 & 25.94 & -88.53 & 1.97 & 0.42 & 0.10 & 12.61 & 4.15 & 25.84 \\
0.4 - 0.5 & 0.090 & 0.000 & 14.82 & 7.26 & 40.58 & -120.04 & 6.31 & 1.48 & 0.65 & 8.51 & 5.78 & 39.94 \\
0.5 - 0.5 & 0.043 & 0.022 & 25.94 & 7.47 & 46.12 & -153.92 & 3.18 & 0.66 & 0.17 & 22.77 & 6.82 & 45.95 \\
0.4 - 0.55 & 0.102 & 0.000 & 15.58 & 8.02 & 50.35 & -139.92 & 7.65 & 1.93 & 0.79 & 7.93 & 6.09 & 49.56 \\
0.5 - 0.55 & 0.077 & 0.003 & 25.06 & 10.63 & 56.48 & -175.68 & 7.09 & 1.58 & 0.63 & 17.97 & 9.05 & 55.85 \\
0.55 - 0.55 & 0.046 & 0.023 & 33.41 & 9.29 & 59.42 & -196.19 & 3.92 & 0.79 & 0.22 & 29.48 & 8.50 & 59.20 \\
0.4 - 0.6 & 0.119 & 0.000 & 15.48 & 8.62 & 63.53 & -163.25 & 10.61 & 2.72 & 1.38 & 4.86 & 5.91 & 62.15 \\
0.5 - 0.6 & 0.100 & 0.000 & 25.21 & 12.41 & 68.81 & -199.98 & 9.12 & 2.27 & 0.85 & 16.09 & 10.14 & 67.96 \\
\enddata
\label{tab:output2}
\end{deluxetable*}
}

{\setlength{\tabcolsep}{2pt}
\begin{deluxetable*}{ccccccccccccc}
\tablecolumns{13}
\tablewidth{0pc}
\tabletypesize{\scriptsize}
\tablecaption{Table of merger remnant properties, continued.}
\tablehead{Simulation & $(T_\mathrm{c})_8$ & $(T_\mathrm{max})_8$ & $\varpi(T_\mathrm{max})$ & $(\rho(T_\mathrm{max}))_6$ & $(S(T_\mathrm{max}))_8$ & $M_\mathrm{enc}(T_\mathrm{max})$ & $(T^\mathrm{z}_\mathrm{max})_8$ & $(\rho(T^\mathrm{z}_\mathrm{max}))_6$ & $M_\mathrm{enc}(T^\mathrm{z}_\mathrm{max})$ & $M_\mathrm{enc}(\frac{1}{4}E_\mathrm{th})$ & $M_\mathrm{enc}(\frac{1}{2}E_\mathrm{th})$ & $M_\mathrm{enc}(\frac{3}{4}E_\mathrm{th})$}
\startdata
0.4 - 0.4 & 1.29 & 1.48 & 0.0003 & 0.66 & 1.19 & 0.000 & 1.95 & 0.23 & 0.116 & 0.199 & 0.400 & 0.608 \\
0.4 - 0.5 & 0.55 & 1.74 & 0.0092 & 0.43 & 1.36 & 0.424 & 2.01 & 0.08 & 0.501 & 0.403 & 0.542 & 0.681 \\
0.5 - 0.5 & 1.86 & 2.03 & 0.0005 & 1.09 & 1.20 & 0.000 & 2.71 & 0.39 & 0.125 & 0.243 & 0.497 & 0.761 \\
0.4 - 0.55 & 0.37 & 1.97 & 0.0090 & 0.44 & 1.41 & 0.484 & 2.35 & 0.11 & 0.531 & 0.471 & 0.597 & 0.723 \\
0.5 - 0.55 & 1.03 & 1.58 & 0.0101 & 0.53 & 1.26 & 0.577 & 2.58 & 0.24 & 0.418 & 0.366 & 0.602 & 0.807 \\
0.55 - 0.55 & 1.97 & 2.40 & 0.0004 & 1.37 & 1.22 & 0.000 & 3.16 & 0.57 & 0.110 & 0.264 & 0.544 & 0.835 \\
0.4 - 0.6 & 0.39 & 2.47 & 0.0090 & 0.38 & 1.57 & 0.557 & 2.70 & 0.07 & 0.628 & 0.546 & 0.647 & 0.751 \\
0.5 - 0.6 & 0.82 & 2.16 & 0.0083 & 0.76 & 1.31 & 0.517 & 2.77 & 0.16 & 0.583 & 0.463 & 0.657 & 0.838 \\
\enddata
\label{tab:output3}
\end{deluxetable*}
}

{\setlength{\tabcolsep}{2pt}
\begin{deluxetable*}{cccccccccccc}
\tablecolumns{12}
\tablewidth{0pc}
\tabletypesize{\scriptsize}
\tablecaption{Table of merger remnant properties, continued.}
\tablehead{Simulation & $(T^\mathrm{cv}_\mathrm{max})_8$ & $\varpi(T^\mathrm{cv}_\mathrm{max})$ & $(\rho(T^\mathrm{cv}_\mathrm{max}))_6$ & $(S(T^\mathrm{cv}_\mathrm{max}))_8$ & $\Omega_\mathrm{c}$ & $\Omega_\mathrm{max}$ & $\varpi(\Omega_\mathrm{max})$ & $M_\mathrm{enc}(\Omega_\mathrm{max})$ & $M_\mathrm{enc}(\frac{1}{4}E_\mathrm{rot})$ & $M_\mathrm{enc}(\frac{1}{2}E_\mathrm{rot})$ & $M_\mathrm{enc}(\frac{3}{4}E_\mathrm{rot})$}
\startdata
0.4 - 0.4 & 0.96 & 0.0031 & 0.67 & 1.05 & 0.35 & 0.35 & 0.0003 & 0.000 & 0.315 & 0.515 & 0.675 \\
0.4 - 0.5 & 1.74 & 0.0092 & 0.43 & 1.36 & 0.16 & 0.22 & 0.0104 & 0.489 & 0.532 & 0.668 & 0.770 \\
0.5 - 0.5 & 1.24 & 0.0034 & 1.13 & 1.04 & 0.46 & 0.46 & 0.0003 & 0.000 & 0.387 & 0.635 & 0.839 \\
0.4 - 0.55 & 1.97 & 0.0090 & 0.44 & 1.41 & 0.15 & 0.24 & 0.0104 & 0.555 & 0.589 & 0.713 & 0.810 \\
0.5 - 0.55 & 1.58 & 0.0101 & 0.53 & 1.26 & 0.34 & 0.34 & 0.0003 & 0.000 & 0.516 & 0.742 & 0.896 \\
0.55 - 0.55 & 1.37 & 0.0033 & 1.44 & 1.03 & 0.52 & 0.52 & 0.0002 & 0.000 & 0.425 & 0.697 & 0.923 \\
0.4 - 0.6 & 2.47 & 0.0090 & 0.38 & 1.57 & 0.14 & 0.27 & 0.0122 & 0.674 & 0.675 & 0.763 & 0.851 \\
0.5 - 0.6 & 2.16 & 0.0083 & 0.76 & 1.31 & 0.24 & 0.28 & 0.0084 & 0.529 & 0.615 & 0.803 & 0.935 \\
\enddata
\label{tab:output4}
\end{deluxetable*}
}

{\setlength{\tabcolsep}{4pt}
\begin{deluxetable*}{cccccccccccc}
\tablecolumns{12}
\tablewidth{0pc}
\tabletypesize{\scriptsize}
\tablecaption{Table of merger remnant properties, continued.}
\tablehead{Simulation & $(I)_{51}$ & $(I_\mathrm{disk})_{51}$ & $(I_\mathrm{ce})_{51}$ & $(L_\mathrm{disk})_{49}$ & $(L_\mathrm{ce})_{49}$ & $\varpi_\mathrm{disk}$ & $\varpi_\mathrm{md}$ & $(\rho(\varpi_\mathrm{md}))_6$ & $(T(\varpi_\mathrm{md}))_8$ & $\Omega(\varpi_\mathrm{md})$ & $h_\mathrm{z}(\varpi_\mathrm{md})$}
\startdata
0.4 - 0.4 & 3.31 & 2.41 & 0.90 & 6.13 & 14.64 & 0.0219 & 0.0344 & 0.00 & 0.44 & 0.08 & 0.0071 \\
0.4 - 0.5 & 6.40 & 5.93 & 0.47 & 16.29 & 8.83 & 0.0168 & 0.0292 & 0.01 & 0.48 & 0.10 & 0.0058 \\
0.5 - 0.5 & 3.23 & 2.29 & 0.94 & 7.37 & 20.10 & 0.0200 & 0.0303 & 0.00 & 0.58 & 0.11 & 0.0059 \\
0.4 - 0.55 & 7.47 & 7.09 & 0.38 & 19.52 & 7.69 & 0.0154 & 0.0302 & 0.01 & 0.46 & 0.10 & 0.0058 \\
0.5 - 0.55 & 6.47 & 5.84 & 0.63 & 14.77 & 14.88 & 0.0172 & 0.0271 & 0.01 & 0.58 & 0.12 & 0.0051 \\
0.55 - 0.55 & 3.26 & 2.31 & 0.95 & 7.96 & 23.00 & 0.0191 & 0.0280 & 0.01 & 0.67 & 0.13 & 0.0054 \\
0.4 - 0.6 & 9.83 & 9.57 & 0.26 & 24.33 & 4.86 & 0.0129 & 0.0278 & 0.01 & 0.62 & 0.11 & 0.0060 \\
0.5 - 0.6 & 6.52 & 6.01 & 0.51 & 19.44 & 12.68 & 0.0159 & 0.0278 & 0.01 & 0.66 & 0.12 & 0.0056 \\
\enddata
\label{tab:output5}
\end{deluxetable*}
}

\bibliography{AutoBibliography}

\end{document}